\documentclass{article}
\usepackage{amsfonts}
\usepackage{stmaryrd}
\usepackage{amssymb}
\usepackage{euscript}
\usepackage{amsthm}
\usepackage{amsmath}
\usepackage{amscd}
\usepackage{latexsym}
\usepackage{mathrsfs}
\usepackage{color}
\usepackage{dsfont}
\usepackage{bm}
\usepackage{eurosym}
\usepackage{booktabs, subfig}
\usepackage[pdftex]{graphicx}

\newtheorem{definition}{Definition}[part]
\newtheorem{proposition}{Proposition}[part]

\newtheorem{assumption}{Assumption}[part]
\newtheorem{lemma}{Lemma}[part]
\newtheorem{corollary}{Corollary}[part]
\newtheorem{remark}{Remark}[part]

\newcommand{\wh}{\widehat}
\newcommand{\wt}{\widetilde}

\newcommand{\Sdelta}{S}
\newcommand{\Udelta}{U}
\newcommand{\Tdelta}{T}

\newcommand{\vsq}{v^2}

\newcommand{\Noti}{P}

\newcommand{\FS}{{\rm \bf FS}}

\newcommand{\FFs}{{\rm \bf FF}^s}

\newcommand{\FCs}{{\rm \bf FC}^s}

\newcommand{\FCshk}{{\rm \bf FC}^{s,h,\kappa}}

\newcommand{\PSshk}{{\rm \bf PS}^{s,h,\kappa}}
\newcommand{\PSj}{{\rm \bf PS}^j}
\newcommand{\PS}{{\rm \bf PS}}

\newcommand{\Rsf}{R^{s,f}}

\newcommand{\Rhh}{R^h}
\newcommand{\Bhh}{B^h}
\newcommand{\rhh}{r^h}

\newcommand{\rb}{r^{\beta}}

\newcommand{\sph}{\alpha^h}
\newcommand{\sps}{\alpha^s}
\newcommand{\spe}{\alpha^e}
\newcommand{\spc}{\alpha^c}
\newcommand{\spu}{\alpha^u}

\newcommand{\wtVfp}{\widetilde{V}^{p,h}}
\newcommand{\wtVf}{\widetilde{V}^{h}}
\newcommand{\wtCh}{\widetilde{C}^{h}}

\newcommand{\Swap}{\FS^{s,h,\kappa}}
\newcommand{\Swapb}{\FS^{s,\beta,\kappa}}
\newcommand{\Swapn}{\FS^{s,h,\kappa}}

\newcommand{\cMf}{\mathcal{M}^f}

\newcommand{\xxx}{X}
\newcommand{\I}{\mathds{1}}

\newcommand{\Q}{\mathbb Q}
\newcommand{\Qf}{\mathbb{Q}^f}
\newcommand{\Qhf}{\widehat{\mathbb{Q}}^f}

\newcommand{\EP}{{\mathbb E}_{\mathbb P}}

\newcommand{\EQf}{{\mathbb E}_{{\mathbb{Q}^f}}}
\newcommand{\EQhf}{{\mathbb E}_{\Qhf}}

\newcommand{\Bb}{B^\beta}

\newcommand{\wtVb}{\widetilde{V}^{\beta}}

\newcommand{\bpp}{{\mathbb P}}
\newcommand{\brr}{{\mathbb R}}
\newcommand{\bff}{{\mathbb F}}

\newcommand{\cF}{\mathcal F}
\newcommand{\cM}{\mathcal M}
\newcommand{\cP}{\mathcal P}

\title{{\Large \bf  PRICING AND HEDGING OF SOFR DERIVATIVES}\vskip 75 pt }

\author{Matthew Bickersteth$\,^{a}$, Yining Ding$\,^{a}$ and Marek Rutkowski$\,^{a,b}$ \\ \\ \\ \\
$^{a\,}$School of Mathematics and Statistics, University of Sydney \\ Sydney, NSW 2006, Australia \\ \\
$^{b\,}$Faculty of Mathematics and Information Science, Warsaw University of Technology \\ 00-661 Warszawa, Poland \\ }

\date{\vskip 25 pt 31 December 2021; Revised on 15 March 2025 \vskip 20 pt}

\begin{document}

\maketitle

\begin{abstract}
The LIBOR has served since the 1970s as a fundamental measure for floating term rates across multiple currencies and maturities. However, in 2017 the Financial Conduct Authority announced the discontinuation of LIBOR from the end of 2021 and the New York Fed declared the Treasury repo financing rate, called the Secured Overnight Financing Rate (SOFR), as a candidate for a new reference rate for interest rate swaps denominated in U.S. dollars. We examine arbitrage-free pricing and hedging of swaps referencing SOFR without and with collateral backing. As hedging instruments, we take SOFR futures and idiosyncratic funding rates for the hedge and margin account. For simplicity, a one-factor model based on Vasicek's equation is used to specify the joint dynamics of several overnight interest rates, including the SOFR and unsecured funding rate.
\end{abstract}

\vskip 40 pt
\noindent KEYWORDS \\
LIBOR, SOFR, SOFR futures, interest rate swap, overnight indexed swap, caplet, swaption
\vskip 4 pt
\noindent JEL CLASSIFICATION  \\
C02, C60, E43, G12, G13


\newpage



\section{Introduction}  \label{sec0}

One of the most important over-the-counter (OTC) derivatives used to manage cash flows are interest rate swaps, which provide an institution the ability to hedge their exposure to interest rate movements in either a single or multiple currencies. There are two key classes of swap contracts, an \emph{interest rate swap} (IRS) in a single currency and a \emph{cross-currency basis swap} (CCBS) in multiple currencies. An \emph{interest rate swap} is an agreement between two parties to exchange the interest on a notional principal in a single currency where, in the most common case of a fixed-for-floating swap, one of the parties to a contract pays a fixed rate, and the other party pays a reference floating rate. In a \emph{cross-currency swap} (CCBS), the initial and terminal exchange of notional principal amounts, the floating interest rates and collateral amounts are based on multiple currencies and thus a CCBS has a more complex structure than an IRS (see, e.g., Ding et al. \cite{DLR2024} or Fujii et al. \cite{FST2010}).

Traditionally, of the most important concepts associated with interest rate swaps has been the London Interbank Offered Rate (LIBOR), which was introduced in 1984 and aimed to reflect the forward-looking cost of interbank borrowing and lending. For almost forty years, LIBOR was ubiquitous as the key benchmark rate for business loans and interest rate swaps worldwide. Although financial instruments benchmarked on LIBOR have been traded in high volumes on the global financial market, in July 2017 the Financial Conduct Authority (FCA) announced the gradual discontinuation of LIBOR. One of the main concerns surrounding the LIBOR was market manipulating activities by some members of the panel. Recall that the LIBOR was a forward-looking term rate based on a survey of a panel of twenty international banks as to what rate they could borrow funds for each currency and maturity but was not directly related to actual trades.

On June 22, 2017, the Alternative Reference Rate Committee (ARRC) announced the Treasury repo financing rate, formally called the Secured Overnight Financing Rate (SOFR), would be a more robust {\it risk-free rate} (RFR) to replace the LIBOR. The SOFR is a backward-looking measure of the overnight repurchasing rate for Treasury securities and it has been published by the New York Federal Reserve since 2018.  As a useful extension of the overnight SOFR, the {\it SOFR Averages} were introduced as either compound or simple averages of the SOFR over rolling 30-, 90-, and 180-calendar day periods.
The {\it SOFR Index} allows for the calculation of compounded average SOFR over custom time periods.
By their construction, the SOFR Averages are manifestly backward-looking term rates, as opposed to forward-looking term rates, such as the LIBOR.

The clear benefit from transition from LIBOR to SOFR is that the latter is determined by market transactions, rather
than by taking a survey, and thus it is considerably less susceptible to banks attempting to manipulate
the market. Additionally, while the LIBOR referenced unsecured borrowing and hence reflected also the interbank credit risk (see, e.g., Filipovi\'c and Trolle \cite{FiT2013}), the repurchase transactions are backed by the Treasury securities as collateral and thus the SOFR is a secured rate with negligible credit risk  since collateral backing protects the lender in the event of default. Since its introduction, the SOFR has gradually became an influential interest rate that banks use to price U.S. dollar-denominated loans and fixed income derivatives. An analogous transition from IBOR to a risk-free rate (RFR) has been recently implemented in other major economies, although some other RFRs are unsecured, that is, based on transactions with no collateral backing.

Although SOFR has many advantages compared to LIBOR,  it should be noted that there are several practical and theoretical issues regarding the transition. One of the primary challenges associated with an RFR, such as the SOFR, is that it has a single reference period (typically, overnight) in contrast to LIBOR that has a range of tenors (typically, from overnight till 12-Month). Since in the case of SOFR there are no directly observable long-term maturities, the forward interest rates need to be inferred from the overnight rate and derivative instruments (e.g. futures contracts). Another key issue, specifically concerning swap contracts, is that the SOFR Average is a backward-looking compound overnight rate over a specific past period (say, one month or three months) while LIBOR has been manifestly a forward-looking term rate. In particular, in an interest rate swap over the period $[\Udelta,\Tdelta]$, the SOFR fixing is not observed until time $\Tdelta$ whereas the LIBOR fixing in the classical IRS was already known at time $T$ (although this was not the case for LIBOR-in-arrears swap with fixing at $\Tdelta$).


Finally, it is worth mentioning that the SOFR was also chosen to replace the \textit{Effective Federal Funds Rate} (EFFR) as the {\it Price Alignment Interest} (PAI). The PAI is the overnight cost of funding collateral, which is debited from the receiver and transferred to the payer to cover the loss of interest on posted collateral (\textit{variation margin}) in centrally cleared transactions. For convenience, the PAI for centrally cleared trades is henceforth called the \textit{collateral rate}, which is the term commonly used for collateralized OTC trades, so the terminology used in this work does not distinguish between centrally cleared and OTC collateralized trades.

The related issue of idiosyncratic discounting in the risk-neutral pricing formula was examined by Piterbarg \cite{P2010} who studied pricing of collateralized contracts on assets (e.g., collateralized stock options) under divergent interest rates, $r_R, r_F$ and $r_C$ for secured (repo) funding, unsecured funding, and collateral, respectively. One of his conclusions was that it suffices to focus on the collateral rate $r_C$ when pricing contracts with theoretical full collateral. Intuitively, the rate on collateral may be viewed as an equivalent of a funding rate when dealing with fully collateralized contracts. Piterbarg's results on funding costs and hedging of a collateralized contract were extended to more general linear and nonlinear setups by several authors, including Bielecki et al. \cite{BCR2018,BR2015}, Brigo et al. \cite{BBFPR2022},  Brigo and Pallavicini \cite{BP2014}, Cr\'epey \cite{C2015a,C2015b}, Fujii et al. \cite{FST2010}, Fujii and Takahashi \cite{FT2013}, Gnoatto and Seiffert \cite{GS2021b}, Nie and Rutkowski \cite{NR2018}, and Pallavicini et al. \cite{PPB2012}.


Our main goal is to examine multi-curve arbitrage-free pricing of collateralized derivatives referencing SOFR through self-financing hedging strategies based on idiosyncratic funding costs and liquidly traded contracts, such as SOFR futures and OIS based on SOFR. The main results furnish closed-form expressions for replicating strategies under the assumption that the factor process $x$ is governed by Vasicek's dynamics and bases are deterministic. A possible extensions to more general dynamics of a factor process (e.g., a multi-factor affine term structure model; see Duffie and Kan \cite{DK1996}, Christensen et al. \cite{CDR2011} or Klingler and Syrstad \cite{KS2021}) is also briefly discussed in Remark \ref{remxx}.  For modeling  issues regarding the dynamics of overnight rates and related econometric studies, we refer to recent works by Alfeus et al. \cite{AGS2017}, Backwell et al. \cite{BMSS2021}, Gellert and Schl\"ogl \cite{GS2021}, Haitfield and Park \cite{HP2019} and Klingler and Syrstad \cite{KS2021}.

In a model-free study, Haitfield and Park \cite{HP2019} show that SOFR futures-implied term rates
accurately predict realized compounded overnight rates during most periods and thus can be used as indicative forward-looking term rates derived from end-of-day SOFR futures prices. In that way, one can obtain synthetic term rates analogous to the term LIBOR rates, which were commonly used in loan contracts. The CME Term SOFR Reference Rates benchmark is a daily set of forward-looking interest rate estimates, calculated and published for 1-month, 3-month, 6-month and 12-month tenors. Fixings of term SOFR are based on transaction data from thirteen consecutive 1-month SOFR futures contracts and five consecutive 3-month CME SOFR futures contracts. Klingler and Syrstad \cite{KS2021} conduct a detailed econometric analysis of overnight and term spreads and provide statistical tests of hypotheses regarding the origin of volatility and spikes observed in those spreads (see also Appendix C in \cite{KS2021} where a stylized model of interbank lending with asymmetric regulatory constraints is examined).




Our research is largely motivated by the recent papers by Mercurio \cite{M2018} and Lyashenko and Mercurio \cite{LM2019a,LM2019b,LM2019c}. Mercurio \cite{M2018} provides an outline for the joint modeling of SOFR/OIS rates and considers the pricing of the respective SOFR futures contracts. His work is further expanded in \cite{LM2019a,LM2019b} where the classical LIBOR Market Model is generalized to cover also the backward-looking rates, such as the SOFR. One of the key concepts in arbitrage-free pricing is the assumption that future cash flows from a derivative security can be replicated by a self-financing trading strategy. Therefore, the present paper attempts to provide greater context to \cite{LM2019a,LM2019b} by showing that SOFR derivatives can be replicated by dynamic trading strategies based on SOFR futures, with idiosyncratic funding costs and possibly also a proportional collateral backing.
It is fair to acknowledge that some pricing formulae obtain in the present paper were already derived in previous works
(see, in particular, Andersen and Bang \cite{AB2020}, Henrard \cite{H2018,H2022}, and Lyashenko and Mercurio \cite{LM2019a,LM2019b}) through other means without making explicit reference to hedging strategies based on SOFR futures.
In contrast, we develop a strict pricing method and we show that they can also be derived in a multi-curve framework for collateralized contracts using mathematically sound replication-based arguments.

The structure of this work is as follows. Section \ref{sec2} gives an overview of overnight rates and the respective definitions for a multi-curve market. In contrast to a single-curve modeling approach, there is no longer a single risk-free interest rate in the market but we deal instead with a family of interest rates corresponding to different segments of interest rate market. We introduce four key rates for consideration: SOFR, EFFR, the unsecured funding rate, and the collateral rate. Having defined the overnight rates, we propose in Section \ref{sec3} a framework for the modeling of the joint dynamics of these interest rates under Vasicek's model. In particular, we obtain a closed-form expression for the SOFR futures. For ease of explicit computations, we focus here on Vasicek's dynamics for the factor process but it is worth noting that Gellert and Schl\"ogl \cite{GS2021} use an extended HJM approach, while Backwell et al. \cite{BMSS2021} choose an affine term structure model to represent the dynamics of SOFR (see also the recent papers by Brace et al. \cite{BGS2024},  Fontana \cite{F2022}, Fontana et al. \cite{FGGS2020}, and Henrard \cite{H2022} for more general models).
In Section \ref{sec3.3}, we deal with arbitrage-free pricing and hedging for swaps referencing SOFR. We also give explicit representations for the arbitrage-free prices of SOFR caps and swaptions. The case of different funding rates available to firms for hedging purposes, in addition to factoring in the level of collateralization, is also considered. For an exposition
of analogous results within the single-curve term structure setup (in particular, the LIBOR and swap rate models), we refer to the monographs by Brigo and Mercurio \cite{BM2001} and Musiela and Rutkowski \cite{MR2005}. We conclude the paper by presenting in Section \ref{sec5} numerical results on pricing and hedging of various SOFR derivatives. We first verify by Monte Carlo simulations the validity of our pricing and hedging results for SOFR swaps, caps and swaptions and we provide a detailed sensitivity analysis.

\section{Multiple Overnight Interest Rates} \label{sec2}

Single-curve term structure models rely on a handy postulate of existence of a unique \textit{risk-free interest rate}, which is available at all times to all market participants. In the classical single short rate modeling approach, it is common to denote by $r_t$ the theoretical instantaneous risk-free rate such that any entity is entitled to a short-term borrowing or lending of cash at rate $r_t$ at time $t$. Then there exists a unique money market account $B$ growing at the rate $r$, as well as a zero-coupon bond price $B(t,T)$ for any maturity $T>0$. All market prices are then driven by the short term rate process $r$, which is usually assumed to enjoy the Markov property. It is also postulated that for every $T$, the bond maturing at time $T$ is a traded asset, which is default-free, and the bond market is arbitrage-free in a suitable sense. All these assumptions are mathematically convenient but it is clear that they are far from reality of fixed income markets.

In an extended multi-curve term structure model, we depart from the simplistic setup where models are built around the assumption of a single instantaneous risk-free rate $r$, which is available to all market participants for borrowing and lending at any time $t \in [0,T^*]$ where $T^*>0$ is the horizon date for the economy. Instead, we model various interest rates representing different segments of financial markets and sources of funds. In contrast to the classical approach, access to the various markets may be restricted and every market is not always available to certain market participants. Additionally, the issues of \textit{credit risk} and \textit{roll-over risk} need also to be examined as, for instance, in Cuchiero et al. \cite{CFG2019}, Fontana et al. \cite{FGGS2020}, Grbac et al. \cite{GPSS2015}, and Mercurio \cite{M2010,M2018}.

The aim of this section is to outline a particular multi-curve framework, which consists of multiple overnight interest rates and related futures contracts.  We start by introducing the notation and assumptions regarding overnight rates in the multi-curve market model, namely, the SOFR, EFFR, an unsecured funding rate, and a collateral rate. It is fairly common in the existing literature to formally associate each short-term rate in a multi-curve setup with a corresponding fictitious zero-coupon bond, which is not assumed to be a traded asset. Rather, our goal is to focus directly on traded interest rate derivatives, such as SOFR futures, fixed-for-floating swaps referencing SOFR and overnight indexed swaps, as opposed to the bond market. Unlike in the classical approach, where the bond and swap markets are considered to be a single market, we acknowledge that the two markets are distinct sectors of the global fixed income market and hence each of them may require a distinct modeling approach.

\subsection{Secured Overnight Financing Rate} \label{sec2.1}


The \textit{repo market} for a given security refers to the sale of an asset between two parties, with
the agreement to repurchase the asset at a later date (often overnight), typically for a higher price.
The repo agreement (i.e., the repurchase agreement) may therefore be considered as a short term loan with the collateral backing of the underlying asset, where the corresponding rate of return upon repurchase is called the
\textit{repo rate}. In particular, the daily SOFR settings are based on the overnight repurchase market where
the underlying assets used as collateral are Treasury securities. More specifically, the daily SOFR setting is computed as
transaction-value weighted median interest rate on overnight U.S. Treasury general (as opposed to {\it special})
collateral repurchase transactions. The SOFR is widely recognized that it represents a broad measure for the cost of borrowing cash overnight collateralized by Treasury securities.

We denote by $\rho^s(t_j)$ the daily SOFR for overnight transactions negotiated on business day $t_j$ and published on the next good business day after transaction day. The \textit{SOFR Averages} are calculated through compounding of the overnight rates over a reference period. Let $n_c$ (resp. $n_b$) denote the number of calendar days (resp. business days) during the period $[\Udelta,\Tdelta]$.
Then the compound \textit{SOFR Average} for the period $[\Udelta,\Tdelta]$ is given by the conventional formula
\begin{align*}
R^s(\Udelta,\Tdelta) := \frac{360}{n_c}\left[\prod^{n_b}_{j=1} \left( 1+\frac{n_j \rho^s(t_j)}{360} \right) - 1 \right]
\end{align*}
where $n_j$ is the number of calendar days to which $\rho^s(t_j)$ applies, and $n_j=1$ for every business day, except for Friday
where $n_j=3$ and Saturday where $n_j=2$. This means that the (compound) SOFR Average employs daily compounding on each business day,
and for any day that is not a business day, the simple interest applies at a rate of interest equal to the SOFR fixing for the preceding business day.
Furthermore, in accordance with broader U.S. dollar money market convention, interest is calculated using the actual number of calendar days, but assuming a 360-day year. The simple \textit{SOFR Average} for the period $[\Udelta,\Tdelta]$ is given by the following expression
\begin{align} \label{sofr}
\bar{R}^s(\Udelta,\Tdelta) := \frac{360}{n_c} \sum^{n_b}_{j=1} \frac{n_j \rho^s(t_j)}{360}.
\end{align}

The compound SOFR Averages are published with tenors of 30-, 90-, and 180-calendar days, that is, for $n_c$ equal to 30, 90, and 180.
For more details on calculations of averages for the SOFR and other overnight rates, see \cite{FRBNY,FSB2019}. We also refer to Lyashenko and Nie \cite{LN2021} for a study of discretization of continuous-time models for backward-looking overnight rates. In a continuous-time framework, we adopt the following definition of the instantaneous SOFR and the compound SOFR Average. Since the simple SOFR Average is not used in what follows, its continuous-time specification is not included in Definition \ref{def2.1}.

\begin{definition} {\rm \label{def2.1}
We denote by $r^s$ the \textit{instantaneous SOFR} so that the continuously compounded SOFR account $B^s$ over $[\Udelta,\Tdelta]$ satisfies
\begin{align} \label{eq2.1}
\frac{B^s_{\Tdelta}}{B^s_\Udelta}= \exp \left(\int_\Udelta^{\Tdelta} r^s_u\, du \right) \approx \prod^{n_b}_{j=1} \left( 1+\frac{n_j \rho^s(t_j)}{360} \right).
\end{align}
The (compound) \textit{SOFR Average} over the period $[\Udelta,\Tdelta]$ is an $\cF_{\Tdelta}$-measurable random variable given by
\begin{align}  \label{eq2.2}
R^s(\Udelta,\Tdelta):=\frac{1}{\delta} \left(\exp \left( \int_\Udelta^{\Tdelta} r^s_u \,du \right) - 1 \right)=\frac{1}{\delta}\left(\frac{B^s_{\Tdelta}}{B^s_{\Udelta}} - 1 \right).
\end{align}
} \end{definition}

It is clear that the compound SOFR for the period $[\Udelta,\Tdelta]$ may only be observed at time $\Tdelta$, which again emphasizes the difference
between a forward-looking LIBOR and a backward-looking SOFR. We henceforth assume the instantaneous SOFR is driven by the underlying \emph{factor process} $x$ supplemented by a basis $\sps$, so that the equality $r^s = x + \sps $ holds.

\subsection{SOFR Futures} \label{sec2.6}

In 2018, the Chicago Mercantile Exchange (CME) introduced SOFR futures contracts referencing One-Month and Three-Month SOFR Averages.  Trading in Eurodollar futures referencing Three-Month LIBOR was discontinued in mid-2023 and hence the role of Eurodollar futures was taken over by futures referencing Three-Month SOFR Average. Each Three-Month SOFR futures contract expires by cash settlement and the CME applies compounding of daily SOFR values between quarterly IMM dates (the third Wednesday of every March, June, September, and December) to determine settlement prices. The method for calculation is exactly the same as for the compound SOFR Average presented in Section \ref{sec2.1}. By convention, the final settlement price for an expiring Three-Month SOFR futures contract equals 100 minus the SOFR rate, compounded daily over the contract’s reference quarter and, like previously Eurodollar futures, Three-Month SOFR futures are sized at \$25 per basis point per annum of contract interest. Quarterly contracts are listed for 39 consecutive quarters and the nearest 6 serial contract months.


Without further mention, we assume throughout that we are given a probability space $(\Omega,\cF,\bff,\bpp)$ where the filtration $\bff$ satisfies the usual conditions of right-continuity and $\bpp$-completeness, and the initial $\sigma$-field $\cF_0$ is trivial. All processes introduced in what follows are assumed to be $\bff$-adapted.

Consistently with equation \eqref{eq2.2}, we find it convenient to adopt the convention of continuously compounded SOFR Average in Definition \ref{def2.6} of the SOFR futures price. The definition and financial interpretation of a {\it martingale measure} $\Qf$ and hence the rationale for equation \eqref{eq2.12} are given in Section \ref{sec3}.

\begin{definition} {\rm \label{def2.6}
A \textit{SOFR futures} contract for the period $[\Udelta,\Tdelta]$ is defined as a futures contract referencing SOFR Average over $[\Udelta,\Tdelta]$ and thus the \textit{SOFR futures rate} is given by, for all $t \in [0,\Tdelta]$,
\begin{align} \label{eq2.12}
\Rsf_t(\Udelta,\Tdelta):=\EQf \big( R^s(\Udelta,\Tdelta) \mid \cF_t \big).
\end{align}
} \end{definition}

In view of \eqref{eq2.2} and \eqref{eq2.12}, the process $\Rsf (\Udelta,\Tdelta)$ is a martingale under $\Qf$ and the following equality holds, for every $t\in [0,\Udelta]$,
\begin{align} \label{eq2.13}
1 + \delta \Rsf_t(\Udelta,\Tdelta)=\EQf \left( e^{\int_\Udelta^{\Tdelta} r^s_u \, du}\,\big|\, \cF_t \right)
\end{align}
and, for every $t \in [\Udelta,\Tdelta]$,
\begin{align} \label{eq2.14}
1 + \delta \Rsf_t(\Udelta,\Tdelta)= e^{\int^t_\Udelta r^s_u\,du} \, \EQf\left( e^{\int_t^{\Tdelta} r^s_u \,du} \,\big|\, \cF_t \right).
\end{align}

We denote by  $f^{s}_t(\Udelta,\Tdelta)$ the SOFR futures price at time $t$ so that $f^{s}_t(\Udelta,\Tdelta)=1-\Rsf_t(\Udelta,\Tdelta)$ and thus the two processes have the same dynamics with respect to $t$ (up to the sign) since  $df^{s}_t(\Udelta,\Tdelta) = -d\Rsf_t(\Udelta,\Tdelta)$. For that reason, it suffices to examine the dynamics of the SOFR futures rate $\Rsf_t(\Udelta,\Tdelta)$.

\subsection{Effective Federal Funds Rate} \label{sec2.2}

\textit{Federal funds} are excess reserves that commercial banks and other financial institutions deposit at regional Federal Reserve banks. These funds can be lent, then, to other market participants with insufficient cash on hand to meet their lending and reserve needs. The \textit{Federal Funds Rate} (aka Fed Funds rate) is the interest rate at which depository institutions are expected to lend reserve balances to other depository institutions overnight on an uncollateralized basis. Since 2008, the Federal Reserve no longer defines a target rate but rather a target range (e.g., 4.75\% to 5.00\%, as of October 2024). In fact, the Federal Open Market Committee (FOMC) cannot force banks to charge the Federal Funds Rate but rather sets a target range as a guidepost. The actual interest rate a lending bank will charge is determined through negotiations between the two banks and the weighted average of interest rates across all transactions of this type is known as the \textit{Effective Federal Funds Rate} (EFFR). The EFFR fixing is based on unsecured transactions between commercial banks, borrowing and lending excess reserves in their exchange account with Federal Reserve banks.
The EFFR fixing published on business day $t_j$, which we denote as $\rho^e(t_j)$, is calculated using the same methodology as used for the SOFR fixing. Similar to the SOFR Average, the \textit{EFFR Average} for the period $[\Udelta,\Tdelta]$ becomes fully known at time $\Tdelta$ and is given by equation \eqref{sofr} with the superscript $s$ replaced by $e$ (see \cite{FRBNY}).
Analogous to Definition \ref{def2.1}, we choose to assume that EFFR can be modeled by an instantaneous interest rate, denoted as $r^e$, as we will later postulate that $r^e=x+\spe$ for a factor process $x$.

\begin{definition} {\rm \label{def2.2}
Let $r^e$ denote the \textit{instantaneous EFFR} so that the continuously compounded EFFR account $B^e$ over $[\Udelta,\Tdelta ]$ satisfies \eqref{eq2.1} with the superscript $s$ replaced by $e$.
The (compound) \textit{EFFR Average} over the period $[\Udelta,\Tdelta]$ is an $\cF_{\Tdelta}$-measurable random variable given by \eqref{eq2.1} with the superscript $s$ substituted with $e$.
} \end{definition}

\subsection{Unsecured Funding Rate} \label{sec2.3}

Commercial banks in the U.S. have two primary ways to borrow money for their short-term operating needs. They can borrow
and loan money to other banks without the need for any collateral using the market-driven interbank rate. The interbank lending system is short-term, typically overnight, and rarely more than a week. They can also borrow funds for their short-term operating requirements from the Federal Reserve Bank. The \textit{discount rate} is the interest rate charged to commercial banks and other financial institutions for short-term loans they take from the Federal Reserve Bank.

We introduce the formal concept of an \textit{unsecured funding rate} for loans with no collateral backing. As an unsecured rate is uncollateralized, there exists a credit risk component in the event of default. In contrast to the previously defined overnight rates,  an unsecured funding rate is not market wide, but rather firm-specific and thus it is assumed that different banks will be able to borrow money at different funding rates. Banks with a high credit rating will be able to borrow funds at a lower rate than banks with a poor credit rating. Hence the next definition should be interpreted in the context of a particular financial institution, rather than the market as a whole.

\begin{definition} {\rm \label{def2.3}
Let $r^u=x+\spu$ denote the instantaneous \textit{unsecured funding rate,} which is firm-specific. The respective continuously
compounded account $B^u$ satisfies $dB^u_t=r^u_tB^u_t\,dt$ with $B^u_0=1$.
} \end{definition}

\subsection{Price Alignment Interest and Collateral Rate} \label{sec2.4}

Another key concept when considering the riskiness of a contract is the collateral backing. Collateralization (also known as the \textit{margining}) refers to a party depositing capital (typically, Treasury notes or bonds) as backing to offset liabilities in case of default. The \textit{collateral rate} is the interest rate paid on collateral (\textit{variation margin}) account by the collateral receiver to the pledging party, that is, the collateral provider.  Although cash is considered the safest form of collateral, many other assets may be deposited as collateral.
Often, when an entity borrows funds to purchase an asset, the asset may be deposited as collateral.
In the case of centrally cleared trades, the collateral rate is called the {\it Price Alignment Interest} (PAI) and the SOFR was recently chosen to replace the EFFR as standard PAI. In an OTC trade, the collateral rate negotiated in the Credit Support Annex (CSA) does not have a specific market definition. Rather, the remuneration rate on collateral is agreed upon by the two parties at the inception of the trade. Therefore, we find it convenient to introduce a generic instantaneous interest rate, henceforth denoted as $r^c$, to represent the interest on margin account and to consider alternative specifications for the generic process $r^c$. For instance, we may set $r^s,\,r^e$ or $r^u$ to play the role of $r^c$, depending on the trade's CSA and prevailing market conventions. Hence we adopt the following generic definition of collateral rate.

\begin{definition} {\rm \label{def2.4}
Let $r^c=x+\spc$ denote the instantaneous \textit{collateral rate}, that is, the remuneration rate on the margin account representing the pledged or received collateral. The respective continuously compounded account $B^c$ satisfies $dB^c_t=r^c_tB^c_t\,dt$ with $B^c_0=1$.
} \end{definition}

\section{Trading Strategies with SOFR Futures}   \label{sec3} 

When considering the arbitrage-free pricing of derivative securities, the price of the contract will vary based, in particular, on the choice of a short rate used for discounting. Under the traditional LIBOR models the risk-free rate $r$ was implemented for discounting and the same rate $r$ was also the short rate corresponding to LIBOR and zero-coupon bond prices. However, post GFC, it has been clear that this simplistic approach is not reflective of the realities of financial markets. In addition, most fixed income derivatives, such as interest rate swaps, are now subject to collateralization and, even more importantly, they also started to use some overnight rate as a variable reference rate for a floating leg. Finally, the question of funding costs for hedging strategies has become much more relevant than in pre GFC framework where a unique short-term rate was identified with the funding rate for a hedge.

\subsection{Futures Trading Strategies for Collateralized Contracts}  \label{sec3.1}

We assume that the underlying probability space $(\Omega , \mathcal{F}, \mathbb{F},\mathbb{P})$ is given and the filtration $\mathbb{F}$ satisfies the usual conditions of right-continuity and $\mathbb{P}$-completeness. Let a semimartingale $f=(f^1,f^2,\dots ,f^m)$ represent a given family of futures prices. A \textit{futures trading strategy} (or, simply, \textit{futures portfolio}) is a process $\varphi=(\varphi^0,\varphi^1,\dots , \varphi^m)$ where the processes $\varphi^k, k=0,1,\dots ,m$ are $\bff$-adapted. Consider a collateral process $C$ where $C_t>0$ means that the collateral amount is held at time $t$ by a bank and can be used for trading (that is, we work here under the fairly common postulate of rehypothecation) whereas $C_t<0$ means that a bank is the collateral pledger at time $t$. 
 Recall that $r^c$ is the interest rate paid on the margin account so that the remuneration of the collateral amount $C_t$ over infinitesimal time period $[t,t+dt]$ is given by $-r_t^c C_t\, dt$, as seen from the bank's perspective.

We henceforth denote by $\Bhh$ the process satisfying $d\Bhh_t=\rhh_t \Bhh_t\, dt$ with $\Bhh_0=1$ where the $\bff$-adapted process $\rhh$ represents the \emph{hedge funding rate}. Hence $\Bhh$ can be any of the processes $B^s, B^e$ or $B^u$, their combination or some other stochastic process, depending on a particular manner in which the cash component of a hedge for a trade under consideration is funded by a bank. Then we have the following definition of
a self-financing collateralized futures portfolio. Notice that equality \eqref{eq3.1} relies on the well known feature of futures contract that they have null value at any time $t$, whereas equality \eqref{eq3.2} can be easily justified by considering first a discrete-time setup or examining the case of piecewise constant hedge ratios $\varphi^{i}$ for $i=1,2,\dots ,m$.  For more comprehensive versions of self-financing conditions \eqref{eq3.1}--\eqref{eq3.2}, the interested reader is referred to, e.g., \cite{BCR2018,BR2015,BBFPR2022,BP2014,NR2018}.

\begin{definition} {\rm \label{def3.1}
The {\it portfolio's value process} $V^p(\varphi )$ of a futures trading strategy $\varphi$ with collateral $C$ equals, for every $t\in [0,T]$,
\begin{align} \label{eq3.1}
V^p_t(\varphi )=\varphi^0_t \Bhh_t
\end{align}
and we say that a futures portfolio $\varphi$ is {\it self-financing} if, for every $t \in [0,T]$,
\begin{align} \label{eq3.2}
V^p_t(\varphi)=V_0(\varphi )+\int_0^t\varphi^0_u\,d\Bhh_u+\sum_{i=1}^m \int_0^t \varphi^{i}_u\,df^i_u+C_t-\int_0^t r^c_u C_u\,du.
\end{align}
The hedger's {\it wealth process} is given by $V_t(\varphi)=V^p_t(\varphi)-C_t$ for every $t\in [0,T]$.
} \end{definition}

For simplicity, the notions of the initial and variation margin for futures contracts are not introduced in Definition \ref{def3.1} and the counterparty credit risk is disregarded throughout this work.  Let us also observe that in the present setup it is not possible to use zero-coupon bonds associated with either the SOFR or the EFFR since they are not traded assets, although their fictitious prices are frequently introduced in papers on SOFR derivatives to facilitate the presentation of formal mathematical computations (see Section \ref{sec3.2}). Finally, since SOFR futures are only traded up to five years, replication of longer term swaps and options would require to roll expired futures into new ones. However, this would introduce some level of incompleteness to our model and hence one would need to deal with a particular method of pricing and hedging under incompleteness.

In the next result, we denote $\wtVf(\varphi):=(\Bhh)^{-1}V(\varphi),\wtVfp(\varphi):=(\Bhh)^{-1}V^p(\varphi)$ and $\wtCh:=(\Bhh)^{-1}C$.

\begin{proposition} \label{pro3.1}
The discounted wealth $\wtVf(\varphi):=(\Bhh)^{-1}V(\varphi)$ of any self-financing futures portfolio $\varphi$ with
a collateral process $C$ satisfies, for every $t\in [0,T]$,
\begin{align} \label{eq3.3}
\wtVf_t(\varphi)=\wtVf_0(\varphi )+\sum_{i=1}^m\int_0^t \varphi^{i}_u(\Bhh_u)^{-1}\,df^i_u
  +\int_0^t (\rhh_u-r^c_u)\wtCh_u\,du.
\end{align}
\end{proposition}

\proof The It\^o integration by parts formula gives
\begin{align*}
d\wtVf_t(\varphi )&=(\Bhh_t)^{-1}\,dV_t(\varphi)+V_t(\varphi)\,d(\Bhh_t)^{-1} \\
&=(\Bhh_t)^{-1}\,d(V^p_t(\varphi)-C_t)+ (V^p_t(\varphi)-C_t)\,d(\Bhh_t)^{-1} \\
&=(\Bhh_t)^{-1}\bigg(\varphi^0_t\,\rhh_t \Bhh_t\,dt+\sum_{i=1}^m\varphi^{i}_t\,df^i_t+dC_t-r^c_t C_t\,dt\bigg)
    -(\Bhh_t)^{-1}\,dC_t- \wtVf_t(\varphi)\,\rhh_t \,dt \\
&=\rhh_t\wtVfp_t(\varphi)\,dt+\sum_{i=1}^m\varphi^{i}_t(\Bhh_t)^{-1}\,df^i_t-r^c_t\wtCh_t\,dt- \rhh_t\wtVfp_t(\varphi)\,dt
    +\rhh_t \wtCh_t \,dt \\
&=\sum_{i=1}^m\varphi^{i}_t(\Bhh_t)^{-1}\,df^i_t+(\rhh_t-r^c_t)\wtCh_t\,dt ,
\end{align*}
which shows that \eqref{eq3.3} is satisfied.
\endproof

\begin{remark} {\rm \label{rem3.1}
It is clear that if the equality $r^c=\rhh$ holds, then
\begin{align} \label{eq3.4}
\wtVf_t(\varphi )=\wtVf_0(\varphi )+\sum_{i=1}^m \int_0^t \varphi^{i}_v (\Bhh_u)^{-1}\,df^i_u
\end{align}
where we now have $\wtVf(\varphi )=(\Bhh)^{-1}V(\varphi )=(B^c)^{-1}V(\varphi)$ and thus collateralization with any process $C$ does not affect the wealth dynamics of a self-financing futures strategy. Needless to say, the choice of collateral $C$ is still crucial for specification of the close-out payment, which is used to settle the trade in the occurrence of the counterparty's default. For related issues regarding explicit modeling of the counterparty credit risk, the interested reader is referred to, e.g., Bielecki et al. \cite{BR2015,BCR2018}, Brigo et al. \cite{BBFPR2022},
Brigo and Pallavicini \cite{BP2014}, Cr\'epey \cite{C2015a,C2015b}, and Pallavicini et al. \cite{PPB2012}.
} \end{remark}

The level of collateralization is usually chosen to be proportional to the current mark-to-market value of a trade, which is here formally represented by the wealth process of a hedging strategy. Therefore, we henceforth assume that $C = -\beta V(\varphi)$ where $\beta$ is an $\bff$-adapted, nonnegative stochastic process reflecting the level of proportionality. Notice that there is no need to assume that the process $\beta$ is a universal process applied to all trades between a bank and its numerous counterparties (or even to various trades with the same counterparty) but it is instead explicitly specified in the Credit Support Annex (CSA), which is a binding document that gives the terms for the provision of collateral by the parties in a trade. Therefore, $\beta$ depends on the CSA complementing a particular trade and thus we make allowance for an arbitrary level of collateralization for a derivative security studied in our market model. In that sense, the level of collateralization can be seen as an additional parameter affecting the contract's price, meaning that one can get a whole spectrum of potential prices for any particular contract within a common market model. The following proposition will allow us to effectively deal with the case of proportional collateralization for an arbitrary derivative security.

Let us denote $r^\beta_t:=(1-\beta_t)\rhh_t+ \beta_t r^c_t$ and $d\Bb_t=r^\beta_t \Bb_t\, dt$ with $\Bb_0=1$.
Representation \eqref{eq3.7} is a very handy tool since we may now perform discounting in the case of the proportional collateralization using the \textit{effective hedge funding rate} $r^\beta$.  The special case an uncollateralized trade is obtained by taking $\beta\equiv 0$ and hence $r^\beta = \rhh$, whereas the case of full collateralization corresponds to $\beta \equiv 1$ so that $r^\beta = r^c$.

\begin{proposition} \label{pro3.2}
The discounted wealth $\wtVb(\varphi):= (\Bb)^{-1} V(\varphi)$ of any self-financing futures portfolio $\varphi$ with the
proportional collateral $C= -\beta V(\varphi)$ satisfies
\begin{align}  \label{eq3.7}
\wtVb_t(\varphi)=\wtVb_0(\varphi)+\sum^m_{i=1}\int_0^t\varphi_u^i (\Bb_u)^{-1}\, df^i_u.
\end{align}
\end{proposition}

\proof
We have $V^p=V+C= V-\beta V = (1-\beta )V$ and thus
\begin{align*}
d\wtVb_t (\varphi)&=(\Bb_t)^{-1}dV_t(\varphi)+V_t(\varphi)\,d(\Bb_t)^{-1} \\
&=(\Bb_t)^{-1}\Big(\varphi_t^0\,d\Bhh_t + \sum^m_{i=1}\varphi_t^i\,df^i_t + \beta_t r^c_t V_t(\varphi)\,dt  \Big)-\rb_t (\Bb_t)^{-1}V_t(\varphi)\,dt \\
&=(\Bb_t)^{-1}\Big(\rhh_t V^p_t(\varphi)\,dt+\sum^m_{i=1}\varphi_t^i\,df^i_t + \beta_t r^c_t V_t(\varphi)\,dt-\rb_t V_t(\varphi)\, dt \Big) \\
&=(\Bb_t)^{-1}\Big((1-\beta_t)\rhh_t V_t(\varphi)+\beta_t r^c_t V_t (\varphi)-\rb_t V_t(\varphi)\Big)dt +\sum^m_{i=1}\varphi_t^i (\Bb_t)^{-1}df^i_t \\
&=\sum^m_{i=1}\varphi_t^i(\Bb_t)^{-1}df^i_t
\end{align*}
and thus \eqref{eq3.7} holds.
\endproof

We are now ready to examine the arbitrage-free properties of the futures market for collateralized trades. To this end, we introduce the following definition.

\begin{definition} {\rm \label{def3.2a}
A probability measure $\Qf$ equivalent to $\bpp$ on $(\Omega, \cF_{T^*})$ is said to be an \textit{equivalent local martingale measure for} $f$ (or, briefly, an ELMM for $f$) if the process $f$ is a $\Qf$-local martingale with respect to $\bff$. We denote by $\cP(f)$ the class of martingale measures for $f$.}
\end{definition}

In the next definition, we consider the class of all trades collateralized at an arbitrary rate $\beta$ and thus the process $\beta$ is not predetermined and may vary from trade to trade. In principle, the choice of a process used for discounting is here arbitrary and thus any strictly positive semimartingale can be used for this purpose, as hence considered to be a \textit{unit}. However, we find it convenient to use the process $\Bb$, which depends not only on collateralization rate $\beta$ but also on the choice of rates $r^h$ and $r^c$ used for hedge funding costs and remuneration of margin account.

\begin{definition} {\rm \label{def3.2}
Given the market model $\cMf = \cM^{f,s,e,u,c,h}=(f,B^s,B^e,B^u,B^c,B^h)$, we say that $\Q$ is an \textit{ELMM for any trade collateralized at rate} $\beta$ in $\cMf$ if $\Q$ is equivalent to $\bpp$ on $(\Omega, \cF_{T^*})$ and the discounted wealth $\wtVb(\varphi)$ is a $\Q$-local martingale with respect to $\bff$ for any self-financing futures portfolio $\varphi$.
} \end{definition}

In view of Proposition \ref{pro3.2} we have the following result, which holds when the variation margin is assumed to be proportional to the wealth process.

\begin{proposition}\label{pro3.3}
If $C= -\beta V(\varphi)$ for an arbitrary process $\beta$, then $\Q$ is a martingale measure for a trade collateralized at the rate $\beta$ in $\cMf$ if and only if $\Q$ is a martingale measure for $f$.
\end{proposition}

In view of Proposition \ref{pro3.3}, it is clear that we may consider trades with divergent collateralization rates and still employ the same martingale measure $\Qf \in \cP(f)$ combined with an idiosyncratic process $\Bb$ used for discounting of cash flows. Hence if we denote by $\cP(\cMf)$ the class of all martingale measures for collateralized contracts in the market model $\cMf$, then the equality $\cP(f)=\cP(\cMf)$ holds. We henceforth work under the assumption that  $C= -\beta V(\varphi)$ and a martingale measure $\Qf$ for $f$ exists and thus we may formulate a minor extension of the classical result on arbitrage-free property of a market model.

\begin{proposition}\label{pro3.4}
If a martingale measure $\Qf$ for $f$ exists and we only allow for trading strategies with a nonnegative wealth, then the market model $\cMf=\cM^{f,s,e,u,c,h}$ is arbitrage-free with respect to any collateral rate $\beta $ assuming that the margin account is proportional to the wealth process.
\end{proposition}

We have the following immediate corollary to Proposition \ref{pro3.4}.

\begin{corollary} \label{cor3.2}
If a contingent claim $X$ maturing at time $\Tdelta$ can be replicated in $\cMf$ through a collateralized strategy with rate $\beta$,
then its arbitrage-free price at any date $t \in [0,\Tdelta ]$ equals
\begin{align}  \label{eq3.8}
\pi^{f,\beta }_t(X)&=\Bb_t\,\EQf \Big((\Bb_{\Tdelta})^{-1} X \,\big|\,\cF_t\Big).
\end{align}
In particular, if $\beta \equiv 0$, then $\pi^{f,\beta }(X)=\pi^{f,h}(X)$ where
\begin{align}   \label{eq3.9}
\pi^{f,h}_t(X)=\Bhh_t\,\EQf \Big((\Bhh_{\Tdelta})^{-1} X \,\big|\,\cF_t\Big).
\end{align}
\end{corollary}

\subsection{Dynamics of the SOFR Futures Price}           \label{sec3.2}

The key traded instrument for hedging of contracts referencing SOFR are SOFR futures contracts. As already mentioned, SOFR futures are traded in high volume and thus create a liquid market for trading. Several papers, including Gellert and Schl\"ogl \cite{GS2021} and Klingler and Syrstad \cite{KS2021} have focused on building the SOFR interest rate curve through stripping the discount factors from SOFR futures. In contrast to those papers, we are mainly concerned with SOFR futures strategies for the purpose of hedging swaps based on SOFR Average and other SOFR-linked derivatives in a multi-curve setup.

We henceforth work under the following standing assumption about the factor process $x$. It should be noticed, however, that our results can be extended to multi-factor models with stochastic bases $\sps ,\spe , \sph$ and $\spc$. For various applications of affine framework, we refer to papers by Christensen et al. \cite{CDR2011}, Cuchiero et al. \cite{CFG2019}, Duffie and Kan \cite{DK1996}, Duffie et al. \cite{DFS2003}, Fontana et al. \cite{FGGS2020}, and Grbac et al. \cite{GPSS2015} (see also Section 4.2 in Brigo and Mercurio \cite{BM2001}).

\begin{assumption} \label{ass3.1}
The factor process $x$ is governed by Vasicek's dynamics, for every $t \in\mathbb{R}_+$,
\begin{align} \label{eq3.10}
dx_t = (a - b x_t)\, dt + \sigma \, dW_t^{\Qf}
\end{align}
where $a,b$ and $\sigma$ are positive constants and $W^{\Qf}$ is a Brownian motion under $\Qf$. The basis terms $\sps, \spe$ and $\spu$ are assumed to be deterministic. Furthermore, we postulate that $\rhh=x+\sph$
where the basis $\sph$ is deterministic.
\end{assumption}

For mathematical convenience, we define the price $B^x(t,\Sdelta)$ of the fictitious bond by postulating that,
for any fixed $\Sdelta>0$ and every $t\in [0,\Sdelta]$,
\[
B^x(t,\Sdelta) := \EQf \bigg( e^{-\int_t^\Sdelta x_u\,du }\,\Big|\,\cF_t \bigg)
\]
but we stress that the bond $B^x(t,\Sdelta)$ is not a traded asset. Furthermore, we set
$B^x(t,\Sdelta)=e^{\int_t^\Sdelta x_u\,du}$ for every $t \geq \Sdelta.$
The following lemma for the classical Vasicek model is well known.

\begin{lemma}\label{lem1.1a}
For any fixed $\Sdelta>0$, the process $B^x(t,\Sdelta)$ satisfies, for all $t\in [0,\Sdelta]$,
\begin{align} \label{eq1.2}
B^x(t,\Sdelta)= e^{m(t,\Sdelta) - n(t,\Sdelta) x_t}
\end{align}
where
\begin{align} \label{eq1.3}
m(t,\Sdelta) = \frac{\sigma^2}{2} \int_t^\Sdelta n^2 (u,\Sdelta)\, du-a \int_t^\Sdelta n(u,\Sdelta)\,du
\end{align}
and
\begin{align} \label{eq1.4}
n(t,\Sdelta) =  \frac{1}{b} \left(1-e^{-b(\Sdelta-t)}\right).
\end{align}
The dynamics of $B^x(t,\Sdelta)$ under $\Qf$ are given by
\begin{align} \label{eq1.5}
dB^x(t,\Sdelta) = B^x(t,\Sdelta) \big( x_t\, dt - \sigma n(t,\Sdelta) \, dW^{\Qf}_t \big).
\end{align}
\end{lemma}

Notice that the function $m(t,\Sdelta)$ can also be computed explicitly yielding
\begin{align} \label{eq1.3x}
m(t,\Sdelta) =\frac{4ab-3\sigma^2}{4b^3}+\frac{\sigma^2-2ab}{2b^2}(\Sdelta-t)
+\frac{\sigma^2-ab}{b^3}e^{-b(\Sdelta-t)}-\frac{\sigma^2}{4b^3}e^{-2b(\Sdelta-t)}.
\end{align}

\begin{remark} {\rm \label{remxx}
A much more flexible Gaussian affine model was examined and fitted to the market data by Skov and Skovmand \cite{SS2021}  who focused on the dynamics of SOFR futures by postulating that $r^s_t = c_0 + c_1^{\perp} X_t$ for some $c_0 \in \brr $ and $c_1 \in \brr^d$ where the Markov process $X$ is governed under $\Qf$ by the generalized Vasicek's equation driven by a $d$-dimensional Brownian motion $W^{\Qf}$
\begin{align*}
dX_t=K(\theta-X_t)\,dt+\Sigma\,dW_t^{\Qf}
\end{align*}
for some $d$-dimensional matrices $K$ and $\Sigma$ and a vector $\theta \in \brr^d$. More specifically, they dealt with two- and three-factor Gaussian arbitrage-free Nelson-Siegel model (AFNS model) proposed in Christensen et al. \cite{CDR2011} (see also Dai and Singleton \cite{DS2000}). Notice that their goal was different from ours since they aimed to infer the forward-looking term SOFR rates from SOFR futures data by identifying the convexity correction via an extended Kalman filter, whereas our focus is on general properties of SOFR futures trading strategies under different funding costs and either without and with collateral backing. In particular, Lemma \ref{lem3.1} can be seen a special case of results from Appendix A in \cite{SS2021} and our further results on pricing and hedging of SOFR swaps, caps and swaptions can be, in principle, extended to the AFNS model. However, one would need to address the issue of market incompleteness, for instance, by complementing the model by additional traded assets referencing SOFR.
} \end{remark}

We define $n(t,\Udelta,\Tdelta):=n(t,\Tdelta)-n(t,\Udelta)$ for every $t\in [0,\Udelta]$. Then we denote, for every $t\in [0,\Udelta]$,
\begin{align*}
\Theta (t,\Udelta,\Tdelta):=\int_t^\Udelta \Big( a n(u,\Udelta,\Tdelta)+\frac{1}{2}\sigma^2 n^2(u,\Udelta,\Tdelta)\Big) du
+\int_\Udelta^\Tdelta\Big(an(u,\Tdelta)+\frac{1}{2}\sigma^2 n^2(u,\Tdelta)\Big) du
\end{align*}
and, for every $t\in [\Udelta,\Tdelta]$,
\begin{align*}
\Theta (t,\Tdelta):=\int_{\Udelta}^t x_u\,d u+\int_t^\Tdelta \Big( a n(u,\Tdelta)+\frac{1}{2}\sigma^2 n^2(u,\Tdelta)\Big) du.
\end{align*}
It is worth noting that for $t\in [\Udelta,\Tdelta]$ the term $\Theta (t,\Tdelta)$ is an $\bff$-adapted stochastic process since it involves the integral of $x$ on $[\Udelta,t]$. Finally, we write $A^s(\Udelta,\Tdelta):=e^{-\int_\Udelta^{\Tdelta} \sps_u \, du}$ where $\sps$ is deterministic.

We will now compute the SOFR futures rate, as given by Definition \ref{def2.6}. Given our assumptions about the modeling of SOFR and the underlying rate process, we may derive an explicit expression for the SOFR futures rate and its dynamics. For similar computations, see Section 6 in Mercurio \cite{M2018}.

\begin{lemma} \label{lem3.1}
Under Assumption \ref{ass3.1}, the SOFR futures rate satisfies, for every $t \in [0,\Udelta]$,
\begin{align} \label{eq3.11}
1+\delta\Rsf_t(\Udelta,\Tdelta)=\big(A^s(\Udelta,\Tdelta)\big)^{-1}e^{n(t,\Udelta,\Tdelta)x_t+\Theta (t,\Udelta,\Tdelta)}
\end{align}
and, for every $t \in [\Udelta,\Tdelta]$,
\begin{align} \label{eq3.12}
1+\delta\Rsf_t(\Udelta,\Tdelta)=\big(A^s(\Udelta,\Tdelta)\big)^{-1}e^{n(t,\Tdelta)x_t+\Theta (t,\Tdelta)}.
\end{align}
The dynamics of the SOFR futures rate are, for every $t \in [0,\Udelta]$,
\begin{align} \label{eq3.13}
d\Rsf_t(\Udelta,\Tdelta)=\delta^{-1}\big(1+\delta\Rsf_t(\Udelta,\Tdelta)\big)
n(t,\Udelta,\Tdelta)\sigma\,dW^{\Qf}_t
\end{align}
and, for every $t \in [\Udelta,\Tdelta]$,
\begin{align} \label{eq3.14}
d\Rsf_t(\Udelta,\Tdelta)=\delta^{-1}\big(1+\delta\Rsf_t(\Udelta,\Tdelta)\big)\,n(t,\Tdelta)\sigma\,dW^{\Qf}_t.
\end{align}
The dynamics of the futures price $f^{s}_t(\Udelta,\Tdelta)=1-\Rsf_t(\Udelta,\Tdelta)$ are,
for every $t \in [0,\Udelta]$,
\begin{align} \label{eq3.14y}
df^s_t(\Udelta,\Tdelta)=\big(f^s_t(\Udelta,\Tdelta)-\delta^{-1} -1\big)\,n(t,\Udelta,\Tdelta)\sigma\,dW^{\Qf}_t
\end{align}
and, for every $t \in [\Udelta,\Tdelta]$,
\begin{align} \label{eq3.14x}
df^s_t(\Udelta,\Tdelta)=\big(f^s_t(\Udelta,\Tdelta)-\delta^{-1} -1\big)\,n(t,\Tdelta)\sigma\,dW^{\Qf}_t.
\end{align}
\end{lemma}

\proof
The process $\Rsf (\Udelta,\Tdelta)$ is a martingale under $\Qf$ and satisfies, for every $t\in [0,\Udelta]$,
\begin{align} \label{eq3.15}
1+\delta \Rsf_t(\Udelta,\Tdelta)=\EQf\Big(e^{\int_\Udelta^{\Tdelta} r^s_u\,du}\,\big|\,\cF_t\Big)=
\big(A^s(\Udelta,\Tdelta)\big)^{-1}\,\EQf\Big(e^{\int_\Udelta^{\Tdelta} x_u\,du}\,\big|\,\cF_t\Big)
\end{align}
and, for every $t \in [\Udelta,\Tdelta]$,
\begin{align} \label{eq3.16}
1+\delta \Rsf_t(\Udelta,\Tdelta)=\big(A^s(\Udelta,\Tdelta)\big)^{-1}\,e^{\int^t_\Udelta x_u\,du}\,
\EQf\Big(e^{\int_t^{\Tdelta}x_u\,du}\,\big|\,\cF_t\Big).
\end{align}
Hence it is clear that it suffices to compute the conditional expectation, for $t\leq \Udelta < \Tdelta$,
\begin{align*}
\EQf\Big(e^{\int_\Udelta^\Tdelta x_u\,du}\,\big|\,\cF_t\Big)=\EQf\Big(e^{\xxx_{\Udelta,\Tdelta}}\,\big|\,\cF_t\Big)
\end{align*}
where $\xxx_{\Udelta,\Tdelta}:=\int_\Udelta^\Tdelta x_u\, du$. It can be checked directly that for all $t<\Udelta$
\begin{align*}
\xxx_{t,\Udelta}:=\int_t^\Udelta x_u\,du=n(t,\Udelta)x_t+\int_t^\Udelta a n(u,\Udelta)\,du +
\int_t^\Udelta \sigma n(u,\Udelta)\,dW^{\Qf}_u
\end{align*}
so that
\begin{align*}
\xxx_{\Udelta,\Tdelta}&=\int_t^\Tdelta x_u\,du-\int_t^\Udelta x_u\,du=\xxx_{t,\Tdelta}-\xxx_{t,\Udelta}  \\
&=\mu(t,\Udelta,\Tdelta)+\int_t^\Tdelta\sigma n(u,\Tdelta)\,dW^{\Qf}_u-\int_t^\Udelta \sigma n(u,\Udelta)\,dW^{\Qf}_u
\end{align*}
where
\begin{align*}
\mu(t,\Udelta,\Tdelta) := n(t,\Udelta,\Tdelta) x_t + \int_t^\Tdelta a n(u,\Tdelta)\, du - \int_t^\Udelta a n(u,\Udelta)\, du .
\end{align*}
Therefore, the conditional distribution with respect to $\cF_t$ under the martingale measure $\Qf$ of the random variable $\xxx_{\Udelta,\Tdelta}$
is Gaussian with the conditional expectation $\mu(t,\Udelta,\Tdelta)$ and the conditional variance satisfying
\begin{align*}
\text{Var}_{\Qf}(\xxx_{\Udelta,\Tdelta} \,|\,\cF_t) = \int_t^\Udelta\sigma^2 n^2(u,\Udelta,\Tdelta)\, du + \int_\Udelta^\Tdelta\sigma^2 n^2(u,\Tdelta)\, du =: v^2(t,\Udelta,\Tdelta).
\end{align*}
Consequently, for all $t \leq \Udelta <\Tdelta$,
\begin{align*}
\EQf \big( e^{\xxx_{\Udelta,\Tdelta}}\,\big|\,\cF_t\big)=e^{\mu(t,\Udelta,\Tdelta)+ \frac{1}{2}v^2(t,\Udelta,\Tdelta)}.
\end{align*}
In view of \eqref{eq3.15} and \eqref{eq3.16}, we conclude that equalities \eqref{eq3.11} and \eqref{eq3.12} are valid.
It is easy to check that the SOFR futures rate is indeed governed by \eqref{eq3.13} and \eqref{eq3.14}
\endproof

\section{Swaps Referencing SOFR Average}    \label{sec3.3}

We are now in a position to introduce the crucial distinctions between the classical single-curve and multi-curve term structure models. Recall that, according to the current market convention regarding the U.S. dollar denominated interest rate swaps, the floating leg payoffs are based on the SOFR Average over each accrual period $[T_{j-1},T_j]$. Hence the exact value of each payoff becomes fully known at the end of the corresponding period and uncertainty about that value gradually decreases during the period, which is a specific feature of a SOFR swap. Recall that in the case of a classical swap referencing LIBOR, the floating leg amount is reset at the beginning of each accrual period and thus a dynamic hedging strategy for the cash flow occurring at time $T_j$ ceases at the beginning of the accrual period and is static between the dates $T_{j-1}$ and $T_{j}$.

We consider the case of a multi-period swap contract with the tenor structure $0 \leq T_0 < T_1 < \dots < T_n$ and we denote $\delta_j = T_j - T_{j-1}$ for $j = 1,2,\dots,n$. Although the basic structure of an interest rate swap referencing SOFR Average is similar to the classical LIBOR swap, it should be noticed that for each accrual period $[T_{j-1},T_j]$ the fixing date $T_j$ of a SOFR swap coincides with the payment date $T_j$ due to the fact that the SOFR Average is backward-looking (in practice, the payment usually occurs one or two good-business days after the last day of the accrual period).

This should be contrasted with the standard LIBOR-in-advance swap where, for each accrual period,  $T_{j-1}$ is the fixing date and $T_j$ is the payment date and thus the cash flow occurring at time $T_j$ is reset at time $T_{j-1}$. Notice also that in a Forward Rate Agreement (FRA), as well as LIBOR-in-arrears swap, the fixing and payment date coincide (although, once again, in practice the payment date may be slightly delayed). Although, in general, the two legs of a swap need not have the same frequency, the basic SOFR swap introduced in Definition \ref{def3.3} enjoys that handy property.

\begin{definition} {\rm \label{def3.3}
At every payment date $T_j, j = 1, 2, \dots, n,$ the net cash flow associated with a payer \textit{forward SOFR swap} equals $\FS^{s,\kappa}_{T_j}(T_{j-1}, T_j) = \big( R^s(T_{j-1}, T_j)-\kappa \big)\delta_j\Noti $. The dates $T_0, T_1, \dots, T_{n-1}$ and $T_1, T_2, \dots, T_{n}$ represent the respective \textit{start dates} and the \textit{payment dates} for the accrual periods. The values $n, \Noti$ and $\kappa$ denote the number of payments (length) of the swap, the notional principal, and the preassigned fixed rate of interest. We will assume, without loss of generality, that $P=1$.
} \end{definition}

It is clear that the unilateral price of a SOFR swap will depend, in particular, on a firm-specific funding rate and a given level of collateralization of a trade. We will first assume that a SOFR swap is uncollateralized and the rate $\rhh$ is used by a bank as a funding rate for a replicating strategy where the instrument used for hedging is the reference SOFR futures. Therefore, we will need to deal with a possible mismatch between the floating leg in a swap, which is based on SOFR Average, and the overnight funding rate $\rhh$, which is used for firm-specific hedging. In the next step, the results from Section \ref{sec3.4} will be extended to the case where a swap is collateralized and the margin account is remunerated at the collateral rate $r^c$.

\subsection{Pricing of Uncollateralized SOFR Swaps}   \label{sec3.4}

Let us assume that a swap is uncollateralized and the short-term rate $\rhh$ is used by a bank as a funding rate when implementing a hedging strategy for a swap. We will examine the replicating strategy in Section \ref{sec3.5} and thus Proposition \ref{pro3.5} and \ref{pro3.6} will be fully substantiated through standard hedging arguments. For the reader's convenience, we first study a single-period SOFR swap referencing the period $[\Udelta,\Tdelta ]$.

\begin{definition} {\rm \label{def3.4}
A payer \textit{single-period SOFR swap} over $[\Udelta,\Tdelta]$, settled in arrears at $\Tdelta$, is a fixed-for-floating swap where at time $\Tdelta$ and per one unit of the nominal value, the long party makes the fixed payment $X_1 =\delta \kappa$ and receives the floating payment $X_2 =\delta R^s(\Udelta,\Tdelta)$ linked to the SOFR Average $R^s(\Udelta,\Tdelta)$ over $[\Udelta,\Tdelta]$. Hence the net cash flow at time $\Tdelta $ equals $\FS^{s,\kappa}_{\Tdelta}(\Udelta,\Tdelta)=\big(R^s(\Udelta,\Tdelta )-\kappa \big)\delta \Noti $ where $\Noti$ is the swap's nominal value.
} \end{definition}

Unlike in the classical case of a swap referencing LIBOR, as the floating leg of a SOFR swap is not known until time $\Tdelta$, the net cash flow of a SOFR swap is not reset at time $T$. Therefore, we need to examine the pricing and hedging of a SOFR swap not only up to time $T$ but also during the accrual period $[\Udelta,\Tdelta]$. Using the martingale measure corresponding to the numeraire $\Bhh$, under the temporary assumption that replication of a swap can be achieved, we also compute the forward SOFR swap rate for an uncollateralized single-period SOFR swap. We stress that all pricing results established in this section will be supported in Section \ref{sec3.5} by replication results (see, in particular, Proposition \ref{pro3.7}).

We denote by $\Swap_t(\Udelta,\Tdelta) $ the price at time $t \in [0,\Tdelta]$  of an uncollateralized SOFR swap with a
fixed rate $\kappa $, which means that $\Swap_t(\Udelta,\Tdelta) = \pi^{f,h}_t(\FS^{s,\kappa}_{\Tdelta}(\Udelta,\Tdelta))$ for all $t\in [0,\Tdelta]$ where $\pi^{f,h}$ denotes the pricing functional given by \eqref{eq3.9}.

\begin{definition} {\rm \label{def3.5}
For any time $t\in [0,\Tdelta]$, the \textit{single-period uncollateralized forward SOFR swap rate} $\kappa^{s,h}_t(\Udelta,\Tdelta)$ is a unique $\cF_t$-measurable random variable representing a fixed rate $\kappa $ such that $\Swap_t(\Udelta,\Tdelta)=0$.
} \end{definition}

Similar to Lyashenko and Mercurio \cite{LM2019a}, we define, for all $t\in [0,\Tdelta]$,
\begin{align*}
\Bhh(t,\Tdelta ):= \EQf \Big(e^{-\int^{\Tdelta}_t \rhh_u \, du}  \,\big|\, \cF_t  \Big) = A^h(t,\Tdelta)B^x(t,\Tdelta).
\end{align*}
where we denote $A^h(t,\Sdelta):=e^{-\int_t^{\Sdelta} \sph_u \, du}$ and $\sph$ is deterministic.
The process $\Bhh(t,\Tdelta )$ can be interpreted as either prices of fictitious (non-traded) zero-coupon bond or, in certain circumstances, prices of traded corporate bond. We also define, for all $t\in [0,\Udelta]$,
\begin{align*}
\Bhh(t,\Udelta) := \EQf \Big(e^{-\int^{\Udelta}_t \rhh_u \, du}  \,\big|\, \cF_t  \Big)
\end{align*}
and we set $\Bhh(t,\Udelta):=e^{\int^t_\Udelta \rhh_u \, du}$ for all $t \in [\Udelta,\Tdelta]$. We denote
\begin{align} \label{zetax}
\wt{\zeta} (t,\Udelta,\Tdelta):=\frac{n(t,\Udelta)}{n(t,\Udelta,\Tdelta)},\quad
\zeta (t,\Udelta,\Tdelta):=\frac{n(t,\Tdelta)}{n(t,\Udelta,\Tdelta)},
\end{align}
Since the dates $\Udelta$ and $\Tdelta$ are fixed for brevity we will also write $\wt{\zeta} (t)= \wt{\zeta}(t,\Udelta,\Tdelta)$ and $\zeta (t)= \zeta (t,\Udelta,\Tdelta)$. The following change of variables result is important since it can be used to express the price and hedging strategy for a contingent claim $X_{\Tdelta}$ directly
in terms of the futures price $\Rsf(\Udelta,\Tdelta)$.

\begin{lemma}\label{lem3.2}
(i) For every $\Sdelta >0$, the process $\Bhh(t,\Sdelta)$ equals, for every $t\in[0,\Sdelta]$,
\begin{align*}
\Bhh(t,\Sdelta)=A^h(t,\Sdelta) \exp \Big(m(t,\Sdelta)-n(t,\Sdelta)x_t\Big)= A^h(t,\Sdelta)B^x(t,\Sdelta)
\end{align*}
and the dynamics of $\Bhh(t,\Sdelta)$ are, for every $t\in[0,\Sdelta]$,
\begin{align*}
d\Bhh(t,\Sdelta)=\Bhh(t,\Sdelta)\left( \rhh_t\,dt -\sigma n(t,\Sdelta)\,dW^{\Qf}_t\right).
\end{align*}
(ii) For every $t \in [0,\Udelta]$, we have
\begin{align*}
\Bhh (t,\Udelta)&=A^h(t,\Udelta)\exp\bigg( m(t,\Udelta)-\wt{\zeta}(t) \Big(\ln \big[A^s(\Udelta,\Tdelta)\big(1+\delta \Rsf_t(\Udelta,\Tdelta)\big)\big] -\Theta (t,\Udelta,\Tdelta)\Big)\bigg), \\
\Bhh (t,\Tdelta)&=A^h(t,\Tdelta)\exp\bigg( m(t,\Tdelta)-\zeta (t) \Big(\ln\big[A^s(\Udelta,\Tdelta)\big(1+\delta \Rsf_t(\Udelta,\Tdelta)\big)\big] -\Theta (t,\Udelta,\Tdelta)\Big)\bigg).
\end{align*}
For every $t\in [\Udelta,\Tdelta]$, we have $\Bhh (t,\Udelta)= \Bhh_t (\Bhh_{\Udelta})^{-1}$ and
\begin{align*}
\Bhh (t,\Tdelta)=A^h(t,\Tdelta)\exp\bigg( m(t,\Tdelta)-\ln\big[A^s(\Udelta,\Tdelta)\big(1+\delta \Rsf_t(\Udelta,\Tdelta)\big)\big]+\Theta (t,\Tdelta)\Big)\bigg)
\end{align*}
or, more explicitly,
\begin{align*}
\Bhh (t,\Tdelta)=\frac{A^{s,h}(t,\Tdelta)B^s_t}{\big(1+\delta \Rsf_t(\Udelta,\Tdelta)\big) B^s_{\Udelta}} \exp \Big(\int_t^{\Tdelta}\sigma^2 n^2(u,\Tdelta)\,du \Big).
\end{align*}
\end{lemma}

\proof
Part (i) is a minor extension of Lemma \ref{lem1.1a} and thus its proof is omitted.
From Lemma \ref{lem3.1}, we obtain, for every $t\in [0,\Udelta]$,
\begin{align}  \label{nUT}
n(t,\Udelta,\Tdelta)x_t=\ln\big[A^s(\Udelta,\Tdelta)\big(1+\delta \Rsf_t(\Udelta,\Tdelta)\big)\big]-\Theta (t,\Udelta,\Tdelta)
\end{align}
and, for every $t\in [\Udelta,\Tdelta]$,
\begin{align} \label{nTT}
n(t,\Tdelta)x_t=\ln\big[A^s(\Udelta,\Tdelta)\big(1+\delta \Rsf_t(\Udelta,\Tdelta)\big)\big]-\Theta (t,\Tdelta).
\end{align}
This shows that part (ii) of the lemma is valid.
\endproof

We first derive the arbitrage-free price of a single-period uncollateralized SOFR swap in terms of fictitious bonds or, equivalently, in view of part (ii) in Lemma \ref{lem3.2}, in terms of the futures price.
Without loss of generality, we assume that the notional principal satisfies $\Noti=1$. Admittedly, similar expressions were obtained in other papers on SOFR swaps where, however, they were not supported by hedging arguments.
Recall that we assume that $\sps$ and $\sph$ are deterministic.

\begin{proposition}\label{pro3.5}
If a single-period uncollateralized SOFR swap can be replicated in $\cMf$, then its arbitrage-free price equals, for all $t \in [0,\Tdelta]$,
\begin{align} \label{eq3.17}
\Swap_t(\Udelta,\Tdelta)= A^{s,h}(\Udelta,\Tdelta) \Bhh(t,\Udelta)-\big( 1 + \delta \kappa\big)  \Bhh(t,\Tdelta)
\end{align}
and thus the single-period uncollateralized forward SOFR swap rate, denoted by $\kappa^{s,h}_t(\Udelta,\Tdelta)$,
equals, for all $t \in [0,\Tdelta]$,
\begin{align} \label{eq3.18}
1 + \delta \kappa^{s,h}_t(\Udelta,\Tdelta)=A^{s,h}(\Udelta,\Tdelta)\frac{\Bhh(t,\Udelta)}{\Bhh(t,\Tdelta)}
= (A^s(\Udelta,\Tdelta))^{-1}\frac{B^x(t,\Udelta)}{B^x(t,\Tdelta)}
\end{align}
where we denote $A^{s,h}(\Udelta,\Tdelta):=e^{\int^{\Tdelta}_{\Udelta}( \sps_u - \sph_u)\, du}$. More explicitly,
for every $t \in [0,\Udelta ]$,
\begin{align}  \label{eq3.18x}
1+\delta \kappa^{s,h}_t(\Udelta,\Tdelta)=\big(1+\delta \Rsf_t(\Udelta,\Tdelta)\big)
\exp\Big(-\int_t^{\Tdelta}\sigma^2 n(u,\Tdelta)n(u,\Udelta,\Tdelta)\,du\Big)
\end{align}
and, for every $t \in [\Udelta,\Tdelta]$,
\begin{align} \label{eq3.18y}
1+\delta \kappa^{s,h}_t(\Udelta,\Tdelta)=\big(1+\delta \Rsf_t(\Udelta,\Tdelta)\big)
\exp\Big(-\int_t^{\Tdelta}\sigma^2 n^2(u,\Tdelta)\,du\Big).
\end{align}
\end{proposition}

\proof
We deduce from Corollary \ref{cor3.2} that the arbitrage-free price of a SOFR swap satisfies
\begin{align*}
\pi^{f,h}_t(\FS^{s,\kappa}_{\Tdelta}(\Udelta,\Tdelta))&=\Bhh_t\,\EQf \Big((\Bhh_{\Tdelta})^{-1}\big(\delta R^s(\Udelta,\Tdelta )-\delta \kappa \big)\,\big|\,\cF_t\Big)\\
&=\Bhh_t\,\EQf \Big((\Bhh_{\Tdelta})^{-1}\big( B^s_{\Tdelta}(B^s_{\Udelta})^{-1}-(1+\delta\kappa)\big)\,\big|\,\cF_t\Big) \\
&=\EQf \Big(e^{-\int^{\Tdelta}_t \rhh_u \,du}\, e^{\int^{\Tdelta}_{\Udelta} r^s_u \,du} \,\big|\, \cF_t  \Big)
    - \big(1 +  \delta \kappa \big)  \Bhh(t,\Tdelta) \\
&=\EQf \Big(e^{-\int^{\Udelta}_t \rhh_u \,du}\,e^{\int^{\Tdelta}_{\Udelta}(r^s_u-\rhh_u)\,du}\,\big|\,\cF_t \Big)
    - \big( 1 +  \delta \kappa  \big)  \Bhh(t,\Tdelta) \\
&=e^{\int^{\Tdelta}_{\Udelta} ( \sps_u - \sph_u) \,du}\Bhh(t,\Udelta) - \big( 1 + \delta \kappa \big)\Bhh(t,\Tdelta) \\
&=A^{s,h}(\Udelta,\Tdelta)\Bhh(t,\Udelta)  - \big( 1 + \delta \kappa\big)  \Bhh(t,\Tdelta)=\Swap_t(\Udelta,\Tdelta).
\end{align*}
Since the forward SOFR swap rate is defined as an $\cF_t$-measurable level of the fixed rate $\kappa$ for which the single-period SOFR swap is valueless at time $t$, it is readily seen that equality \eqref{eq3.18} holds for every $t \in [0,\Tdelta ]$. Equation \eqref{eq3.18x} is a direct consequence of \eqref{eq3.18} and representations
of $\Bhh(t,\Udelta)$ and $\Bhh(t,\Tdelta)$ obtained in part (ii) of Lemma \ref{lem3.2}. Since ${\zeta}(t)-\wt{\zeta}(t)=1$
we have that, for every $t \in [0,\Udelta]$
\begin{align*}
1 + \delta \kappa^{s,h}_t(\Udelta,\Tdelta)=\big(1+\delta \Rsf_t(\Udelta,\Tdelta)\big)\exp\bigg(m(t,\Udelta)-m(t,\Tdelta)+\Theta (t,\Udelta,\Tdelta)\Big)\bigg)
\end{align*}
and after simple computations we get \eqref{eq3.18x}. For every $t \in [\Udelta,\Tdelta]$, we obtain
\begin{align*}
1+\delta \kappa^{s,h}_t(\Udelta,\Tdelta)=\big(A^s(\Udelta,\Tdelta)B^x(t,\Tdelta)\big)^{-1}\exp\Big(\int_{\Udelta}^t x_u\,du\Big),
\end{align*}
which yields \eqref{eq3.18y}.
\endproof

Observe that, for every $t \in [0,\Udelta]$, the price $\Swap_t(\Udelta,\Tdelta)$ is a function of the futures price $\Rsf(\Udelta,\Tdelta)$ and,  for every $t \in [\Udelta,\Tdelta]$, it is a function of $\Rsf(\Udelta,\Tdelta)$ and the ratio $B^h_t (B^h_{\Udelta})^{-1}$ or, equivalently, the ratio $B^s_t (B^s_{\Udelta})^{-1}$. An analogous conclusion holds for the forward SOFR swap rate, specifically, at any date $t\in [0,\Tdelta]$ the random variable $\kappa^{s,h}_t$ is a function of the futures price $\Rsf(\Udelta,\Tdelta)$ and a similar property is valid for every $t \in [\Udelta ,\Tdelta]$.  Furthermore, we can formulate an immediate consequence of \eqref{eq3.17} and \eqref{eq3.18}, which holds under the assumptions
of Proposition \ref{pro3.5}.

\begin{corollary} \label{cor3.3}
The price of a single-period uncollateralized SOFR swap with the fixed rate $\kappa$ satisfies, for all $t \in [0,\Tdelta ]$,
\begin{align} \label{eq3.19}
\Swap_t(\Udelta,\Tdelta)=\delta\big(\kappa^{s,h}_t(\Udelta,\Tdelta)-\kappa\big)\Bhh(t,\Tdelta).
\end{align}
In particular, the price at time $t$ of the SOFR swap started at time $0$ with the fixed rate $\kappa $ equal to the forward SOFR swap rate $\kappa^{s,h}_0(\Udelta,\Tdelta)$ equals
\begin{align} \label{eq3.20}
\Swap_t(\Udelta,\Tdelta)=\delta \big(\kappa^{s,h}_t(\Udelta,\Tdelta)-\kappa^{s,h}_0(\Udelta,\Tdelta)\big)\Bhh(t,\Tdelta).
\end{align}
\end{corollary}

Proposition \ref{pro3.5} can be easily extended to a multi-period uncollateralized SOFR swap, as specified in Definition \ref{def3.3}. Let $\Swapn_t(T_0,n)$ denote the (ex-dividend) arbitrage-free price at time $t \in [0,T_0]$ of a multi-period uncollateralized SOFR swap. The \textit{multi-period uncollateralized forward SOFR swap rate} $\kappa^{s,h}_t(T_0,n)$ at time $t\in [0,T_n]$ is a unique $\cF_t$-measurable random variable, which represents a unique level of a fixed rate $\kappa $ for which $\Swapn_t(T_0,n)=0$.

\begin{proposition}\label{pro3.6}
If a multi-period uncollateralized SOFR swap can be replicated in $\cMf$, then its arbitrage-free price equals, for all $t\in [0,T_0]$,
\begin{align} \label{eq3.21}
\Swapn_t(T_0,n)= \sum^n_{j=1} \Big(A^{s,h}(T_{j-1},T_j) \Bhh(t,T_{j-1})- \big( 1 + \delta_j \kappa \big) \Bhh(t,T_j) \Big)
\end{align}
and the multi-period uncollateralized forward SOFR swap rate $\kappa^{s,h}_t(T_0,n)$ equals, for all $t \in [0,T_0]$,
\begin{align} \label{eq3.22}
\kappa^{s,h}_t(T_0,n)= \frac{\sum^n_{j=1} \Big( A^{s,h}(T_{j-1},T_j) \Bhh(t,T_{j-1})  - \Bhh(t,T_j) \Big) }{\sum^n_{j=1}\delta_j \Bhh(t,T_j)}
\end{align}
where
\begin{align*}
A^{s,h}(T_{j-1},T_j):= e^{\int^{T_j}_{T_{j-1}} (\sps_u - \sph_u) \,du}.
\end{align*}
Furthermore, for all $t\in [0,T_0]$,
\begin{align}
\Swapn_t(T_0,n)= \sum^n_{j=1}\delta_j \Bhh(t,T_j) \big(\kappa^{s,h}_t(T_0,n)-\kappa \big).   \label{eq3.23}
\end{align}
For $t \in [T_{k-1}, T_k]$, where $k =1,2,\dots, n-1$, the price of a multi-period uncollateralized SOFR swap $\Swapn_t(T_0,n)$ and the SOFR swap rate $\kappa^{s,h}_t(T_k,n-k)$ are given by expressions analogous to \eqref{eq3.21} and \eqref{eq3.22} with $j=k,k+1,\dots ,n$ and $\Bhh(t,T_{k-1}) = \exp \Big(\int^t_{T_{k-1}} \rhh_u \, du \Big)$.
\end{proposition}

Before stating an alternative pricing result for a SOFR swap, we need to introduce some auxiliary notation.
It is already known from \eqref{eq3.13} and \eqref{eq3.14} that the process $Y:=1+\delta \Rsf(\Udelta,\Tdelta)$ satisfies the linear SDE
\begin{align} \label{3.18a}
dY_t=Y_t\sigma_Y(t,\Udelta,\Tdelta) \, dW^{\Qf}_t
\end{align}
where $\sigma_Y(t,\Udelta,\Tdelta)= \sigma n(t,\Udelta,\Tdelta)$ for all $t \in [0,\Udelta ]$ and $\sigma_Y(t,\Udelta,\Tdelta)=\sigma n(t,\Tdelta)$ for all $t \in [\Udelta,\Tdelta]$.
Hence we have that, for all $t\in [0,\Tdelta ]$,
\begin{align*}
Y_{\Tdelta}=Y_t\exp\bigg(\int_t^{\Tdelta}\sigma_Y(u,\Udelta,\Tdelta)\,dW^{\Qf}_u-\frac{1}{2}\int_t^{\Tdelta}\sigma^2_Y(u,\Udelta,\Tdelta)\,du\bigg)
=Y_t e^{\xi_Y(t,\Udelta,\Tdelta)-\frac{1}{2}v^2_Y(t,\Udelta,\Tdelta)}
\end{align*}
where
\begin{align} \label{3.18b}
\xi_Y(t,\Udelta,\Tdelta):=\int_t^{\Tdelta} \sigma_Y(u,\Udelta,\Tdelta) \, dW^{\Qf}_u,\quad v^2_Y(t,\Udelta,\Tdelta):=\int_t^{\Tdelta}\sigma^2_Y(u,\Udelta,\Tdelta)\,du.
\end{align}
Furthermore, we write $\xi (t,\Tdelta):=\int_t^{\Tdelta}\sigma n(u,\Tdelta)\,dW^{\Qf}_u$ and
\begin{align}  \label{3.18bb}
\xi (t,\Tdelta):=\int_t^{\Tdelta}\sigma n(u,\Tdelta)\,dW^{\Qf}_u,\quad \vsq(t,\Tdelta):=\int_t^{\Tdelta}\sigma^2 n^2(u,\Tdelta)\,du
\end{align}
so that  $\xi_Y(t,\Udelta,\Tdelta)= \xi (t,\Tdelta)$ and $v^2_Y(t,\Udelta,\Tdelta)=\vsq(t,\Tdelta)$ for every
$t\in [\Udelta ,\Tdelta]$.
Finally, we denote $\vsq(t,\Udelta):=\int_t^{\Udelta}\sigma^2 n^2(u,\Udelta)\,du$ for
$t\in [0,\Udelta]$ and $\vsq (t,\Udelta)=0$ for $t \in [\Udelta,\Tdelta]$.

The next result gives an alternative representation for the arbitrage-free price of a SOFR swap in terms of the SOFR futures rate and the factor process $x$, which is convenient when examining SOFR futures options and SOFR caps
(see Proposition \ref{pro3.10}). Recall that $A^h(t,\Tdelta):=e^{-\int_t^{\Tdelta} \sph_u\,du }$.

\begin{proposition}\label{pro3.5a}
If a single-period uncollateralized SOFR swap can be replicated in $\cMf$, then its arbitrage-free price equals, for all $t \in [0,\Tdelta]$,
\begin{align*}
\Swap_t(\Udelta,\Tdelta)= A^h(t,\Tdelta)e^{\rho(x_t)}\Big( Y_t
e^{\frac{1}{2}(\vsq(t,\Udelta )-v^2_Y(t,\Udelta,\Tdelta))}-(1+\delta \kappa) e^{\frac{1}{2} \vsq(t,\Tdelta)}\Big)
\end{align*}
where $Y_t=1+\delta \Rsf_t(\Udelta,\Tdelta)$ and
\begin{align}  \label{3.18nn}
\rho(x_t):=-n(t,\Tdelta)x_t-\int_t^{\Tdelta} a n(u,\Tdelta)\, du .
\end{align}
\end{proposition}

\proof
As in the proof of Proposition \ref{pro3.5}, we observe that, for every $t\in [0,\Tdelta]$,
\begin{align*}
\Swap_t(\Udelta,\Tdelta)& = \Bhh_t \, \EQf \Big( (\Bhh_{\Tdelta})^{-1}\big( R^s(\Udelta,\Tdelta ) -\kappa\big) \delta \,\big|\, \cF_t\Big)\\
& = \EQf \Big(\Rhh(t,\Tdelta)\big(\Rsf_{\Tdelta}(\Udelta,\Tdelta ) -\kappa\big) \delta \,\big|\, \cF_t\Big)\\
& =A^h(t,\Tdelta)\, \EQf \Big(A^x(t,\Tdelta)\big( Y_{\Tdelta}-(1+\delta\kappa)\big)  \,\big|\, \cF_t\Big)
\end{align*}
where $Y_T= 1+\delta \Rsf_T(\Udelta,\Tdelta)$ and
\begin{align*}
\Rhh (t,\Tdelta) = \Bhh_t (\Bhh_{\Tdelta})^{-1}=e^{-\int_t^{\Tdelta} r^h_u\,du } = A^h(t,\Tdelta)  A^x(t,\Tdelta) 
\end{align*}
where $A^x(t,\Tdelta):=e^{-\int_t^{\Tdelta}x_u\,du}$. Furthermore, the dynamics of $x$ are given by \eqref{eq3.10}
and thus
\begin{align*}
\int_t^{\Tdelta}x_u\,du=n(t,\Tdelta)x_t+\int_t^{\Tdelta}an(u,\Tdelta)\,du+\int_t^{\Tdelta}\sigma n(u,\Tdelta)\,dW^{\Qf}_u,
\end{align*}
which in turn implies that $A^x(t,\Tdelta)=e^{\rho(x_t)}e^{-\xi (t,\Tdelta)}$ where $\rho(x_t)$ and $\xi (t,\Tdelta)$
are given by \eqref{3.18nn} and \eqref{3.18bb}, respectively.
Since $\rho (x_t)$ and $Y_t$ are $\cF_t$-measurable, we obtain
\begin{align*}
&\Swap_t(\Udelta,\Tdelta)=A^h(t,\Tdelta)\,\EQf \Big(  Y_{\Tdelta}A^x(t,\Tdelta) - (1+\delta \kappa)A^x(t,\Tdelta) \,\big|\, \cF_t\Big)\\
&=A^h(t,\Tdelta)e^{\rho(x_t)} \,\EQf \Big( Y_t e^{\xi_Y(t,\Udelta,\Tdelta)- \xi (t,\Tdelta)-\frac{1}{2}v^2_Y(t,\Udelta,\Tdelta)}
-(1+\delta \kappa)e^{-\xi (t,\Tdelta)} \,\big|\, \cF_t\Big) \\
& =A^h(t,\Tdelta)e^{\rho(x_t)} \Big( Y_t e^{\frac{1}{2}{\rm Var}(\eta_1)-\frac{1}{2}v^2_Y(t,\Udelta,\Tdelta)}
-(1+\delta \kappa)e^{\frac{1}{2}{\rm Var}(\eta_2)} \Big)
\end{align*}
since the random variable $(\eta_1,\eta_2)$ where $\eta_1 :=\xi_Y(t,\Udelta,\Tdelta)- \xi (t,\Tdelta)$ and $\eta_2:=- \xi (t,\Tdelta)$ is independent of $\cF_t$ and has the Gaussian distribution with
zero mean and the variances ${\rm Var}(\eta_1)=\vsq(t,\Udelta)$ and
${\rm Var}(\eta_2)=\vsq(t,\Tdelta)$ for every $t \in [0,\Tdelta]$.
Hence it is clear that the asserted pricing formula is valid.
\endproof

\begin{remark} {\rm \label{rem4.1x}
From the proof of Lemma \ref{lem3.2} we know that (see \eqref{nUT} and \eqref{nTT}), for every $t\in [0,\Udelta]$,
\begin{align*}
\rho(x_t)=-\zeta (t)\Big(\ln\big[A^s(\Udelta,\Tdelta)\big(1+\delta\Rsf_t(\Udelta,\Tdelta)\big)\big]
 -\Theta (t,\Udelta,\Tdelta)\Big)-a\int_t^{\Tdelta}n(u,\Tdelta)\,du
\end{align*}
and, for every $t\in [\Udelta,\Tdelta]$,
\begin{align*}
\rho(x_t)=-\ln\big[A^s(\Udelta,\Tdelta)\big(1+\delta \Rsf_t(\Udelta,\Tdelta)\big)\big]+\Theta(t,\Tdelta)-a\int_t^{\Tdelta}n(u,\Tdelta)\,du,
\end{align*}
which may be used to show that the pricing formulae from Propositions \ref{pro3.5} and \ref{pro3.5a} are equivalent.}
\end{remark}

\subsection{Hedging of Uncollateralized SOFR Swaps} \label{sec3.5}

To support the pricing formulae from Propositions \ref{pro3.5} and \ref{pro3.5a} through replication,
we will now establish the existence of a replicating strategy for an uncollateralized SOFR swap where the bank uses the rate $\rhh$ for funding of the hedge. In fact, it is easy to show that the considered model is complete so our goal
is to find explicit representations for hedge ratios in terms of market observables, $x_t$ and $\Rsf_t(\Udelta,\Tdelta)$ or,
equivalently, in terms of the SOFR futures rate $\Rsf_t(\Udelta,\Tdelta)$ only.  We will later see that the same feature is valid for hedging strategies for caps and swaptions.

In contrast to the classical single-period swaps referencing LIBOR, the value of the SOFR swap payoff continues to evolve for $t \in [\Udelta,\Tdelta]$ and thus the hedge ratios are no longer static during the accrual period $[\Udelta,\Tdelta]$.
As the replicating strategy is expected to be a continuous process, it is natural to check that $\varphi^1_{\Udelta^-} = \varphi^1_{\Udelta^+}$, which is indeed the case.

Recall that a single-period SOFR swap referencing the period $[\Udelta,\Tdelta]$ has the payoff at $\Tdelta$ given by
\begin{align*}
X=\delta  R^s(\Udelta,\Tdelta)-\delta\kappa = B^s_{\Tdelta}(B^s_{\Udelta})^{-1}- (1+\delta \kappa ). 
\end{align*}
To identify a replicating strategy for an uncollateralized SOFR swap, it suffices to use Proposition \ref{pro3.1} with $C=0$. We formulate a result for a single-period swap but to obtain a replicating strategy for a SOFR swap it suffices to note that a multi-period swap can be viewed as a portfolio of single-period swaps. Recall that $\Bhh (t,\Udelta)$ and $\Bhh (t,\Tdelta)$ are given by Lemma \ref{lem3.2}.

\begin{proposition}\label{pro3.7}
A single-period uncollateralized SOFR swap with a fixed rate $\kappa $ can be replicated by the futures trading strategy $\varphi=(\varphi^0,\varphi^1)$ funded at the rate $\rhh$ where, for all $t \in [0,\Tdelta]$,
\begin{align} \label{eq3.24}
\varphi^0_t \Bhh_t=A^{s,h}(\Udelta,\Tdelta) \Bhh(t,\Udelta) - (1 + \kappa \delta) \Bhh(t,\Tdelta)=\Swap_t(\Udelta,\Tdelta).
\end{align}
Furthermore $f^1=f^s(\Udelta,\Tdelta)$ and $\varphi^1$ equals, for all $t \in [0,\Udelta]$,
\begin{align} \label{eq3.25}
\varphi^1_t=-\frac{(1+\delta\kappa)\Bhh(t,\Tdelta)\zeta(t,\Udelta, \Tdelta)-A^{s,h}(\Udelta,\Tdelta)
\Bhh(t,\Udelta)\wt{\zeta}(t,\Udelta,\Tdelta)}{\delta^{-1}\big(1+\delta\Rsf_t(\Udelta,\Tdelta)\big)}
\end{align}
and, for all $t \in [\Udelta,\Tdelta]$,
\begin{align} \label{eq3.26}
\varphi^1_t =- \frac{(1+\delta\kappa)\Bhh(t,\Tdelta)}{\delta^{-1}(1+\delta \Rsf_t(\Udelta,\Tdelta))}=
- \frac{(1+\delta\kappa)A^{s,h}(t,\Tdelta)B^s_t(B^s_{\Udelta})^{-1}}
 {\delta^{-1}\big(1+\delta \Rsf_t(\Udelta,\Tdelta)\big)^2}\exp\bigg(\int_t^{\Tdelta}\sigma^2 n^2(u,\Tdelta)\,du\bigg).
\end{align}
\end{proposition}

\proof
We wish to find the trading strategy $\varphi$ replicating the swap value so that
\begin{align*}
dV_t(\varphi)=\varphi^0_t\,d\Bhh_t+\varphi^1_t\,df^s_t(\Udelta,\Tdelta)=d\Swap_t(\Udelta,\Tdelta).
\end{align*}
The floating leg of the SOFR swap and the SOFR futures contract are driven by the same Brownian motion and thus we can compute $\varphi^1_t$ by equating the respective diffusion terms. On the one hand, we deduce from equation \eqref{eq3.13} that the dynamics of the futures price are, for all $t\in [0,\Udelta]$,
\begin{align} \label{fut3.22}
df^{s}_t(\Udelta,\Tdelta)=-\delta^{-1} \big(1+\delta\Rsf_t(\Udelta,\Tdelta)\big)n(t,\Udelta,\Tdelta)\sigma\,dW_t^{\Qf}
\end{align}
and thus the wealth process of $\varphi$ satisfies
\begin{align} \label{fut3.22x}
dV_t (\varphi)=\varphi^0_t\,d\Bhh_t-\varphi^1_t\delta^{-1}\big(1+\delta\Rsf_t(\Udelta,\Tdelta)\big)n(t,\Udelta,\Tdelta)
\sigma\,dW_t^{\Qf}.
\end{align}
On the other hand, the dynamics of the price process $\Swap(\Udelta,\Tdelta)$ can be derived by applying It\^{o}'s formula to equation \eqref{eq3.17}, for all $t\in [0,\Udelta]$,
\begin{align*}
d\Swap_t(\Udelta,\Tdelta)&= A^{s,h}(\Udelta,\Tdelta)\,d\Bhh(t,\Udelta)-(1 + \delta\kappa)\,d\Bhh(t,\Tdelta) \\
&= A^{s,h}(\Udelta,\Tdelta)\Bhh(t,\Udelta)\left(\rhh_t\,dt- n(t,\Udelta)\sigma\,dW^{\Qf}_t\right) \\
&-(1+\delta\kappa)\Bhh(t,\Tdelta)\left(\rhh_t\,dt-\sigma n(t,\Tdelta)\sigma\,dW^{\Qf}_t\right) \\
&=\rhh_t\Big( A^{s,h}(\Udelta,\Tdelta)\Bhh(t,\Udelta)-(1+\delta\kappa)\Bhh(t,\Tdelta)\Big)dt \\
&+\Big((1+\delta\kappa)\Bhh(t,\Tdelta)n(t,\Tdelta)-A^{s,h}(\Udelta,\Tdelta)\Bhh(t,\Udelta)n(t,\Udelta)\Big)\sigma \,dW^{\Qf}_t.
\end{align*}
By equating the drift and diffusion terms for any  $t\in [0,\Udelta]$, we obtain equality \eqref{eq3.25}. Notice that the component $\varphi^0_t$ can also be computed from the equalities $V_t (\varphi)=\Swap_t(\Udelta,\Tdelta)=\varphi^0_t\Bhh_t$ for all $t\in [0,\Tdelta]$.

For $t \in [\Udelta,\Tdelta]$, the dynamics of the SOFR futures price are obtained from equation \eqref{eq3.14}
\begin{align*}
df^s_t(\Udelta,\Tdelta)=-\delta^{-1}\big(1+\delta \Rsf_t(\Udelta,\Tdelta)\big)\, n(t,\Tdelta)\sigma\,dW_t^{\Qf}
\end{align*}
so that
\begin{align*}
dV_t (\varphi)=\varphi^0_t\,d\Bhh_t-\varphi^1_t\delta^{-1}\big(1+\delta \Rsf_t(\Udelta,\Tdelta)\big)\,n(t,\Tdelta)\sigma\,dW_t^{\Qf}
\end{align*}
and thus price of SOFR swap satisfies, for all $t\in [\Udelta,\Tdelta ]$,
\begin{align*}
&d\Swap_t(\Udelta,\Tdelta)=A^{s,h}(\Udelta,\Tdelta)\,d\Bhh(t,\Udelta)-(1+\delta\kappa)\,d\Bhh(t,\Tdelta) \\
&=A^{s,h}(\Udelta,\Tdelta)\Bhh(t,\Udelta)\rhh_t\,dt-(1+\delta\kappa)\Bhh(t,\Tdelta)\big(\rhh_t\,dt- n(t,\Tdelta)\sigma \,dW^{\Qf}_t\big).
\end{align*}
As in the first step, we identify the hedge ratios $\varphi^0_t$ and $\varphi^1_t$ by equating the drift and diffusion terms of the respective processes. We obtain the same expression \eqref{eq3.24} for $\varphi^0_t$ as for $t\in [0,\Udelta]$ and we now get equality \eqref{eq3.26} for $\varphi^1_t$. This feature was expected since the dynamics of the SOFR swap price and SOFR futures change when the swap's reference period $[\Udelta,\Tdelta]$ is reached.
\endproof

\subsection{Pricing and Hedging of Collateralized SOFR Swaps}      \label{sec3.6}

In the next step, we study a collateralized SOFR swap when $\rhh$ is used as a funding rate for the hedge.  We henceforth assume that $C = -\beta V(\varphi)$ where $\beta$ is a deterministic function. Before determining the arbitrage-free price of a collateralized SOFR swap, we define the auxiliary process $\Bb(t,\Sdelta)$. Recall that we denote $r^\beta_t:= (1-\beta_t) \rhh_t+ \beta_t r^c_t$ and $d\Bb_t=r^\beta_t \Bb_t\, dt$ with $\Bb_0=1$. We also write $\alpha^{\beta}_t:= (1-\beta_t )\sph_t + \beta_t\spc_t$ so that $r^\beta=x_t+(1-\beta_t)\sph_t+\beta\spc_t=x_t+\alpha_t^{\beta}$.

We now define the fictitious bond price by setting, for every $t\in [0,\Sdelta]$,
\begin{align*}
\Bb(t,\Sdelta):=\EQf \left( e^{ -\int_t^\Sdelta r^{\beta}_u \,du} \,\big|\, \cF_t \right)=A^{\beta}(t,\Sdelta)B^x(t,\Sdelta)
\end{align*}
where $A^{\beta}(t,\Sdelta):=e^{-\int_t^{\Sdelta}\alpha_u^{\beta}\, du}$.
Furthermore, we set, for every $t > \Sdelta$,
\begin{align*}
\Bb (t,\Sdelta) := \exp \left(  \int^t_\Sdelta  r^\beta_u\, du\right)
\end{align*}
and we note that $d\Bb (t,\Sdelta)=r^\beta_t\Bb(t,\Sdelta)\,dt$ for all $t > \Sdelta$. The proof of the next lemma
is obvious.

\begin{lemma}\label{lem3.3}
If $\alpha^{\beta}$ is deterministic, then the process $\Bb(t,\Sdelta)$ satisfies, for all $t \in [0,\Sdelta]$,
\begin{align*}
\Bb(t,\Sdelta)=\exp \bigg(m(t,\Sdelta)-n(t,\Sdelta)x_t-\int_t^\Sdelta \alpha_u^{\beta}\, du\bigg)
\end{align*}
where  for all $t \in [0,\Sdelta]$. The dynamics of $\Bb(t,\Sdelta)$ are
\begin{align*}
d\Bb(t,\Sdelta)=\Bb (t,\Sdelta)\exp\Big(r^\beta_t\,dt-\sigma n(t,\Sdelta)\,dW_t^{\Qf}\Big).
\end{align*}
\end{lemma}

We denote by $\Swapb_t(\Udelta,\Tdelta) $ the price at time $t \in [0,\Tdelta]$  of a single-period collateralized SOFR swap with fixed rate $\kappa $, that is, $\Swapb_t(\Udelta,\Tdelta)=\pi^{f,\beta}_t(\FS^{s,\kappa}_{\Tdelta}(\Udelta,\Tdelta))$ for all $t\in [0,\Tdelta]$ where $\pi^{f,\beta}$ is the pricing functional given by \eqref{eq3.8}. By definition, the \textit{collateralized forward SOFR swap rate} $\kappa^{s,\beta}_t(\Udelta,\Tdelta)$ is the fair rate in a collateralized forward SOFR swap with the collateralization rate $\beta$. The following result extends Proposition \ref{pro3.5} to the case of a collateralized SOFR swap. Finally, we write
\begin{align*}
A^{s,\beta}(\Udelta,\Tdelta):=e^{\int_{\Udelta}^{\Tdelta}(\sps_u-\alpha_u^{\beta})\, du}.
\end{align*}

\begin{proposition}\label{pro3.8}
If $\beta$ is deterministic, then the arbitrage-free price of a single-period collateralized SOFR swap equals, for all $t \in [0,\Tdelta]$,
\begin{align} \label{eq3.27}
\Swapb_t(\Udelta,\Tdelta)=A^{s,\beta}(\Udelta,\Tdelta)\Bb (t,T)-\big(1+\delta \kappa\big)\Bb (t,\Tdelta)
\end{align}
and thus the collateralized forward SOFR swap rate $\kappa^{s,\beta}_t(\Udelta,\Tdelta)$ satisfies, for all $t \in [0,\Tdelta ]$,
\begin{align}  \label{eq3.28}
1+\delta \kappa^{s,\beta}_t(\Udelta,\Tdelta)=A^{s,\beta}(\Udelta,\Tdelta)\,\frac{\Bb(t,T)}{\Bb(t,\Tdelta)}.
\end{align}
\end{proposition}

\proof
As in the proof of Proposition \ref{pro3.5}, by applying \eqref{eq3.8} to the swap's payoff, we obtain
\begin{align*}
\pi^{f,\beta}_t(\FS^{s,\kappa}_{\Tdelta}(\Udelta,\Tdelta))&=\Bb_t\,\EQf \Big((\Bb_{\Tdelta})^{-1}\big(\delta R^s(\Udelta,\Tdelta )-\delta \kappa \big)\,\big|\,\cF_t\Big)\\
&= B_t^\beta\,\EQf\left((\Bb_{\Tdelta})^{-1}\Big( B^s_{\Tdelta}(B^s_{\Udelta})^{-1}-(1+ \delta \kappa )\,\big|\,\cF_t \Big)\right) \\
&=\EQf \left(e^{-\int^{\Udelta}_t r_u^\beta \, du}\,e^{\int^{\Tdelta}_{\Udelta}(r^s_u-r_u^\beta)\,du}\,\big|\,\cF_t\right)
- \big(1+\delta\kappa \big)  B^\beta(t,\Tdelta) \\
&= A^{s,\beta}(\Udelta,\Tdelta)\Bb(t,\Udelta)-(1+\delta\kappa)\Bb(t,\Tdelta)=\Swapb_t(\Udelta,\Tdelta),
\end{align*}
as was required to show.
\endproof

We are now in a position to examine the issue of dynamic replication of a SOFR swap with collateral backing.
Hence the dynamics of the wealth process are now given by Proposition \ref{pro3.1} with $C = -\beta V(\varphi)$.
Then we obtain from Definition \ref{def3.1}
\[
\varphi^0_t \Bhh_t=V^p_t(\varphi)=V_t(\varphi)+ C_t=(1-\beta_t)V_t(\varphi)
\]
and thus the wealth process of a futures trading strategy $\varphi$ under collateralization satisfies
\begin{align*}
V_t(\varphi)&=V_0(\varphi )+\int_0^t\varphi^0_u\Bhh_u \rhh_u\,du+\sum_{i=1}^m\int_0^t \varphi^{i}_u\,df^i_u+\int_0^t r^c_u \beta V_u(\varphi)\,du \\
&= V_0(\varphi)+\int_0^t\varphi^0_u \Bhh_u(1-\beta_u)^{-1}r^\beta_u\,du+\sum_{i=1}^m \int_0^t\varphi^{i}_u\,df^i_u .
\end{align*}
In the case of the full collateralization we have $\beta =1$ and $\varphi_t^0=0$ for every $t\in [0,\Tdelta ]$ since $V^p_t(\varphi )=0$ for all $t\in [0,\Tdelta]$. This means that the initial price and every cash flow generated by trading in futures contracts is immediately posted as collateral and no cash is held in the funding account, which in turn implies that the futures hedge is effectively financed using the rate $r^h=r^c$. Notice that Proposition \ref{pro3.9} can be extended to a multi-period collateralized SOFR swap to obtain a result analogous to Proposition \ref{pro3.6}.

\begin{proposition}\label{pro3.9}
If $\beta$ is deterministic, then a single-period collateralized SOFR swap with $C=-\beta V(\varphi)$ can
be replicated by the SOFR futures strategy $\varphi=(\varphi^0,\varphi^1)$ funded at the rate $\rhh$ where,
for all $t \in [0,\Tdelta]$,
\begin{align*}
\varphi^0_t \Bhh_t = (1-\beta_t )\big( A^{s,\beta}(\Udelta,\Tdelta)\Bb(t,\Udelta)
-(1+\delta\kappa)\Bb(t,\Tdelta)\big)=(1-\beta_t )\Swapb_t(\Udelta,\Tdelta).
\end{align*}
Furthermore $f^1=f^s(\Udelta,\Tdelta)$ and $\varphi^1$ equals, for all $t \in [0,\Udelta]$,
\begin{align*}
\varphi^1_t= - \frac{(1+\delta\kappa)\Bb(t,\Tdelta) \zeta(t,\Udelta,\Tdelta)-A^{s,\beta}(\Udelta,\Tdelta)\Bb(t,\Udelta) \wt{\zeta}(t,\Udelta,\Tdelta)}{\delta^{-1}\big(1+\delta \Rsf_t(\Udelta,\Tdelta)\big)},
\end{align*}
and, for all $t \in [\Udelta,\Tdelta]$,
\begin{align*}
\varphi^1_t=-\frac{(1+\delta\kappa)\Bb(t,\Tdelta)}{\delta^{-1}\big(1+\delta \Rsf_t(\Udelta,\Tdelta)\big)}.
\end{align*}
In particular, for a fully collateralized SOFR swap we have $\beta_t=1$ and thus $\varphi^0_t=0$ for all $t\in [0,\Tdelta]$.
\end{proposition}

\proof
The computations are analogous to the proof of Proposition \ref{pro3.7}.  We have, for all $t \in [0,\Udelta]$,
\begin{align*}
dV_t(\varphi )=\varphi^0_t\Bhh_t(1-\beta_t)^{-1}r^\beta_t\,dt-\varphi^{1}_t\delta^{-1}\big(1+\delta \Rsf_t(\Udelta,\Tdelta)\big)n(t,\Udelta,\Tdelta)\sigma\,dW_t^{\Qf}
\end{align*}
and, for all $t \in [\Udelta,\Tdelta]$,
\begin{align*}
dV_t(\varphi )=\varphi^0_t \Bhh_t(1-\beta_t)^{-1}r^\beta_t\, dt- \varphi^{1}_t\delta^{-1}\big(1+\delta \Rsf_t(\Udelta,\Tdelta)\big)n(t,\Tdelta)\sigma\,dW_t^{\Qf}.
\end{align*}
The arbitrage-free price $\Swapb(\Udelta,\Tdelta)$ of a collateralized SOFR swap with $C=-\beta V(\varphi)$ is given by \eqref{eq3.27} and thus we obtain, for all $t \in [0,\Udelta]$,
\begin{align*}
&d\Swapb_t(\Udelta,\Tdelta)=\delta^{-1}\Big( A^{s,\beta}(\Udelta,\Tdelta) d\Bb(t,\Udelta)-(1+\kappa \delta) d\Bb(t,\Tdelta)\Big) \\
&=\delta^{-1}\Big(A^{s,\beta}(\Udelta,\Tdelta)\Bb(t,\Udelta)-(1+\delta\kappa)\Bb(t,\Tdelta)\Big)r^\beta_t\,dt \\
&+\delta^{-1}\Big((1+\delta\kappa)\Bb(t,\Tdelta)n(t,\Tdelta)- A^{s,\beta}(\Udelta,\Tdelta)\Bb(t,\Udelta)n(t,\Udelta)\Big)\sigma\,dW^{\Qf}_t
\end{align*}
and, for all $t \in [\Udelta,\Tdelta]$,
\begin{align*}
d\Swapb_t(\Udelta,\Tdelta)&=\delta^{-1}\Big(A^{s,\beta}(\Udelta,\Tdelta)\Bb(t,\Udelta)-(1 + \delta\kappa) \Bb(t,\Tdelta)\Big)r^\beta_t\,dt \\
&+\delta^{-1}(1 + \delta\kappa)\Bb(t,\Tdelta)n(t,\Tdelta)\sigma\, dW^{\Qf}_t.
\end{align*}
By equating the drift and diffusion terms, we obtain the hedge ratios $\varphi^0_t$ and $\varphi^1_t$ given in the statement of the proposition.
\endproof

\subsection{Pricing and Hedging of SOFR Futures Options and Caps}   \label{sec3.7}

A \textit{SOFR cap} is a portfolio of European style options where the holder receives the cash difference between the SOFR Average and a pre-determined strike interest rate, provided that the SOFR Average exceeds a pre-determined strike level. Let $\kappa >0$ and $\Noti$ denote the pre-determined strike and notional principal of a cap, respectively. In an interest rate cap referencing LIBOR, the rate considered over the period $[T_{j-1},T_j]$ is the forward LIBOR determined at time $T_{j-1}$ and thus the payoff from the $j$th caplet is known in advance (i.e., at time $T_{j-1}$) since it equals $(L_{T_{j-1}}(T_{j-1}, T_j)-\kappa)^+\delta_j\Noti $ where $L_{T_{j-1}}(T_{j-1}, T_j)$ is the LIBOR at time $T_{j-1}$ referencing the accrual period $[T_{j-1},T_j]$. In contrast, the payoff from the $j$th SOFR caplet equals $(R^s(T_{j-1},T_j)-\kappa)^+ \delta_j\Noti $ and thus it becomes fully known at time $T_j$, although an accurate approximation of the settlement payoff becomes available when time $t$ approaches $T_j$.

\begin{definition} {\rm \label{def3.7}
A \textit{SOFR cap} with strike $\kappa >0$ and maturity $T_n$, which is settled in arrears at dates $T_j,\,j = 1,2, \dots, n$ has at each date $T_j$ the cash flow equal to $\FCs_{T_j}(\kappa)=(R^s(T_{j-1}, T_j)-\kappa)^+\delta_j\Noti $ where $\FCs_{T_j}(\kappa)$ is the payoff from the $j$th SOFR caplet. Similarly, a \textit{SOFR floor} with strike $\kappa>0$ and the same tenor structure has the cash flow at $T_j$ equal to $\FFs_{T_j}(\kappa) = (\kappa- R^s(T_{j-1}, T_j))^+\delta_j\Noti $.
} \end{definition}

As usual, without loss of generality, we set $\Noti = 1$. For the sake of simplicity, in Sections \ref{sec3.7} and \ref{sec3.8}, we focus on the case of uncollateralized caps and swaptions. Using equality \eqref{eq3.9}, we obtain
\begin{align*}
\FCshk_t=\Bhh_t\,\EQf \bigg( \sum^n_{j=1}(\Bhh_{T_j})^{-1}\FCs_{T_j}(\kappa)\,\big|\,\cF_t\bigg)
= \Bhh_t\,\EQf \bigg(\sum^n_{j=1}(\Bhh_{T_j})^{-1}\big( R^s(T_{j-1}, T_j)-\kappa\big)^+ \delta_j\,\big|\,\cF_t\bigg)
\end{align*}
and thus it suffices to compute the arbitrage-free price of the $j$th caplet with the accrual period $[\Udelta,\Tdelta]$ where
we denote $\Udelta=T_{j-1}$ and $\Tdelta = T_{j-1}+\delta_j=T_j$ for any fixed $j \in \{1,2,\dots,n\}$.

Since the equality $\big( R^s(\Udelta, \Tdelta)-\kappa\big)^+=\big( \Rsf_{\Tdelta}(\Udelta, \Tdelta)-\kappa\big)^+$
holds, it is clear that the caplet's payoff can also be interpreted as the payoff of a SOFR futures option maturing
at time $\Tdelta $. Therefore the pricing and hedging results for a SOFR caplet cover also a SOFR futures option.

The next result gives a closed-form expression for the arbitrage-free price of a SOFR caplet or, equivalently, a SOFR futures option. We use here the notation introduced in Section \ref{sec3.4}. In particular,  the dynamics of the process
$Y_t = 1+\delta \Rsf_t(\Udelta,\Tdelta)$ and the variance $v^2_Y(t,\Udelta,\Tdelta)$ are given by \eqref{3.18a} and \eqref{3.18b}, respectively, and the quantities $\rho(x_t), \vsq(t,\Udelta)$ and $\vsq(t,\Tdelta)$ were used in Proposition \ref{pro3.5a}. Recall that $\vsq(t,\Udelta)=0$ for $t \geq U$ and $v^2_Y(t,\Udelta,\Tdelta)=\vsq(t,\Tdelta)$ for $t \geq U$.

\begin{proposition}\label{pro3.10}
The arbitrage-free price of a SOFR caplet with the expiration date $T$, strike $\kappa>0$ and accrual period $[\Udelta,\Tdelta]$ equals, for all $t\in [0,\Udelta]$,
\begin{align*}
\FCshk_t= A^h(t,\Tdelta)e^{\rho(x_t)}\Big( Y_t e^{\frac{1}{2}(\vsq(t,\Udelta)-v^2_Y(t,\Udelta,\Tdelta))} N(h_+(t))
-(1+\delta \kappa) e^{\frac{1}{2} \vsq(t,\Tdelta)}N(h_{-}(t))\Big)
\end{align*}
where
\begin{align*}
h_+(t)=h_+(t,\Udelta,\Tdelta):=\frac{\ln \frac{Y_t}{1+\delta \kappa}+\frac{1}{2}\big(\vsq(t,\Udelta)-\vsq(t,\Tdelta)\big)}{v_Y(t,\Udelta,\Tdelta)}
\end{align*}
and
\begin{align*}
h_-(t)=h_-(t,\Udelta,\Tdelta):=\frac{\ln \frac{Y_t}{1+\delta \kappa}+\frac{1}{2}\big(\vsq(t,\Udelta)-\vsq(t,\Tdelta)-2v^2_Y(t,\Udelta,\Tdelta)\big)}{v_Y(t,\Udelta,\Tdelta)}.
\end{align*}
Furthermore, for every $t\in [\Udelta,\Tdelta]$
\begin{align*}
\FCshk_t= A^h(t,\Tdelta)e^{\rho(x_t)}\Big(Y_t e^{-\frac{1}{2}v^2(t,\Tdelta)} N(h_+(t))
-(1+\delta \kappa) e^{\frac{1}{2} \vsq(t,\Tdelta)}N(h_{-}(t))\Big)
\end{align*}
where
\begin{align*}
h_+(t)=h_+(t,\Udelta,\Tdelta):=\frac{\ln\frac{Y_t}{1+\delta \kappa}- \frac{1}{2}v^2(t,\Tdelta)}{v(t,\Tdelta)}
\end{align*}
and
\begin{align*}
h_-(t)=h_-(t,\Udelta,\Tdelta):=\frac{\ln \frac{Y_t}{1+\delta \kappa}-\frac{3}{2}v^2(t,\Tdelta)}{v(t,\Tdelta)}.
\end{align*}
\end{proposition}

\proof
We use the notation from the proof of Proposition \ref{pro3.5a}. To find the arbitrage-free price of a caplet,
it suffices to evaluate the conditional expectation, for every $t\in [0,\Tdelta]$,
\begin{align*}
\FCshk_t& = \Bhh_t \, \EQf \Big( (\Bhh_{\Tdelta})^{-1}\big( R^s(\Udelta,\Tdelta ) -\kappa\big)^+ \delta \,\big|\, \cF_t\Big)\\
& = \EQf \Big(\Rhh(t,\Tdelta)\big(\Rsf_{\Tdelta}(\Udelta,\Tdelta ) -\kappa\big)^+ \delta \,\big|\, \cF_t\Big)\\
& =A^h(t,\Tdelta)\, \EQf \Big(A^x(t,\Tdelta)\big( Y_{\Tdelta}-(1+\delta\kappa)\big)^+  \,\big|\, \cF_t\Big)
\end{align*}
where $Y_t = 1+\delta \Rsf_t(\Udelta,\Tdelta)$ and
$A^x(t,\Tdelta)=e^{-\int_t^{\Tdelta}x_u\,du}=e^{\rho(x_t)-\xi (t,\Tdelta)}$ for all $t\in [0,\Tdelta ]$.
Since $e^{\rho (x_t)}$ is $\cF_t$-measurable and positive, we have
\begin{align*}
&\FCshk_t=A^h(t,\Tdelta)\,\EQf \bigg( \Big( Y_{\Tdelta}A^x(t,\Tdelta) - (1+\delta \kappa)A^x(t,\Tdelta) \Big)^+ \,\big|\, \cF_t\bigg)\\
&=A^h(t,\Tdelta)e^{\rho(x_t)} \,\EQf \bigg( \Big( Y_t e^{\xi_Y(t,\Udelta,\Tdelta)- \xi (t,\Tdelta)-\frac{1}{2}v^2_Y(t,\Udelta,\Tdelta)}
-(1+\delta \kappa)e^{-\xi (t,\Tdelta)}\Big)^+ \,\big|\, \cF_t\bigg) \\
& =A^h(t,\Tdelta)e^{\rho(x_t)}\,\EQf \bigg( \Big( Y_t e^{\frac{1}{2}{\rm Var}(\eta_1)-\frac{1}{2}v^2_Y(t,\Udelta,\Tdelta)} e^{\widehat{\eta}_1}
-(1+\delta \kappa)e^{\frac{1}{2}{\rm Var}(\eta_2)} e^{\widehat{\eta}_2}\Big)^+ \,\big|\,\cF_t\bigg)
\end{align*}
where we denote $\eta_1 :=\xi_Y(t,\Udelta,\Tdelta)- \xi (t,\Tdelta), \eta_2:=- \xi (t,\Tdelta)$
and $\widehat{\eta}_i :=\eta_i -\frac{1}{2}\,{\rm Var}\,(\eta_i)$ for $i=1,2$. The random variable $(\eta_1,\eta_2)$ is independent of $\cF_t$ and has the Gaussian distribution with zero mean and the variances ${\rm Var}(\eta_1)=\vsq(t,\Udelta)$ and ${\rm Var}(\eta_2)=\vsq(t,\Tdelta)$ for every $t \in [0,\Tdelta]$. To complete the computations, we use Lemma \ref{lem3.4} to obtain
\begin{align*}
\FCshk_t= A^h(t,\Tdelta)e^{\rho(x_t)}\, \Big( Y_t e^{\frac{1}{2}{\rm Var}(\eta_1)-\frac{1}{2}v^2_Y(t,\Udelta,\Tdelta)} N(h_+(t))-(1+\delta \kappa) e^{\frac{1}{2}{\rm Var}(\eta_2)}N(h_{-}(t))\Big)
\end{align*}
where $h_{\pm}(t)=h_{\pm}(t,\Udelta,\Tdelta)=\frac{1}{k} \ln \frac{c_1}{c_2}\pm \frac{1}{2} k,\, k=\sqrt{{\rm Var}(\eta_1-\eta_2)}=v_Y(t,\Udelta,\Tdelta)$ and
\begin{align*}
 \frac{c_1}{c_2} = \frac{Y_t}{1+\delta\kappa} \,e^{\frac{1}{2}{\rm Var}\,(\eta_1) -\frac{1}{2}v^2_Y(t,\Udelta,\Tdelta)-\frac{1}{2}{\rm Var}\,(\eta_2)}.
\end{align*}
It is now easy to check that $h_+(t)$ and $h_-(t)$ are given by the equalities from the statement of the proposition.
\endproof

\begin{lemma}\label{lem3.4}
 Let $(\eta_1,\eta_2 )$ be a zero-mean, jointly Gaussian (non-degenerate), two-dimensional random variable
 on a probability space $(\Omega , \cF,\bpp )$.  Then for arbitrary positive real numbers $c_1$ and $c_2$ we have
\begin{align*}
\EP \Big( c_1 e^{\,\eta_1 -\frac{1}{2}\, {\rm Var}(\eta_1)}-
 c_2 e^{\eta_2-\frac{1}{2}\, {\rm Var}(\eta_2)}\Big)^+=c_1N(h_+)-c_2 N(h_{-})
\end{align*}
 where
\begin{align*}
 h_{\pm}=\frac{1}{k} \ln \frac{c_1}{c_2}\pm \frac{1}{2} k, \quad k=\sqrt{{\rm Var}\, (\eta_1-\eta_2)}.
\end{align*}
\end{lemma}

The hedging strategy for a SOFR caplet (and hence also for a SOFR futures option and a SOFR cap) can be found by differentiating the pricing formula from Proposition \ref{pro3.10} and identifying the hedge ratios $\varphi^0_t$ and $\varphi^1_t$, as was done in the proof of Proposition \ref{pro3.7}. Since our market model is complete, the existence of a replicating strategy for a SOFR caplet (or a SOFR futures option) is obvious. To make the expressions for the hedging strategy more concise and facilitate the computations of hedging strategy, we introduce the auxiliary notation
\begin{align*}
\Gamma_t := A^h(t,\Tdelta)e^{\rho(x_t)} Y_t \,e^{\frac{1}{2}(\vsq(t,\Udelta)-v^2_Y(t,\Udelta,\Tdelta))}
\end{align*}
and
\begin{align*}
\Lambda_t :=  A^h(t,\Tdelta)e^{\rho(x_t)}(1+\delta\kappa)e^{\frac{1}{2}\vsq(t,\Tdelta)}.
\end{align*}
Then the pricing formula for a SOFR caplet established in Proposition \ref{pro3.7} becomes
\begin{align*}
\FCshk_t= \Gamma_t  N\big(h_+(t,\Udelta,\Tdelta)\big) -\Lambda_t N\big(h_{-}(t,\Udelta,\Tdelta)\big).
\end{align*}
Recall that the functions $\wt{\zeta} (t,\Udelta,\Tdelta)$ and  $\zeta (t,\Udelta,\Tdelta)$ are given by \eqref{zetax}.

\begin{proposition}\label{pro3.11}
A SOFR caplet with a fixed strike $\kappa $ can be replicated by a futures trading strategy $\varphi=(\varphi^0,\varphi^1)$
funded at the rate $\rhh$ where, for all $t \in [0,\Tdelta]$,
\begin{align*} 
\varphi^0_t \Bhh_t=A^{s,h}(\Udelta,\Tdelta) \Bhh(t,\Udelta)-(1+\delta\kappa) \Bhh(t,\Tdelta)=\FCshk_t.
\end{align*}
Furthermore $f^1=f^s(\Udelta,\Tdelta)$ and $\varphi^1$ equals, for all $t \in [0,\Udelta]$,
\begin{align*} 
\varphi^1_t=\frac{v_Y(t,\Udelta,\Tdelta)\big[\wt{\zeta}(t,\Udelta,\Tdelta)\Gamma_t N\big(h_{+}(t)\big)-\zeta(t,\Udelta,\Tdelta)\Lambda_t N\big(h_{-}(t)\big)\big]
-\Gamma_t n\big(h_{+}(t)\big)+\Lambda_t n\big(h_{-}(t)\big)}{\delta^{-1}v_Y(t,\Udelta,\Tdelta)\big(1+\delta \Rsf_t(\Udelta,\Tdelta)\big)}
\end{align*}
and, for all $t \in [\Udelta,\Tdelta]$,
\begin{align*} 
\varphi^1_t = \frac{-v_Y(t,\Udelta,\Tdelta)\Lambda_t N\big(h_{-}(t)\big)-\Gamma_t n\big(h_{+}(t)\big)+\Lambda_t n\big(h_{-}(t)\big)}{\delta^{-1}v_Y(t,\Udelta,\Tdelta)\big(1+\delta \Rsf_t(\Udelta,\Tdelta)\big)}.
\end{align*}
\end{proposition}

\proof
We already know from the proof of Proposition \ref{pro3.7} that the dynamics of the wealth of a futures trading strategy $\varphi$ are, for all $t\in [0,\Udelta]$,
\begin{align*}
dV_t (\varphi)=\varphi^0_t\,d\Bhh_t-\varphi^1_t\delta^{-1} \big(1+\delta \Rsf_t(\Udelta,\Tdelta)\big)n(t,\Udelta,\Tdelta)\sigma\,dW_t^{\Qf}
\end{align*}
and, for all $t\in [\Udelta,\Tdelta]$,
\begin{align*}
dV_t (\varphi)=\varphi^0_t\,d\Bhh_t-\varphi^1_t\delta^{-1}\big(1+\delta \Rsf_t(\Udelta,\Tdelta)\big)n(t,\Tdelta)\sigma\,dW_t^{\Qf}.
\end{align*}
Replication of a caplet means that $V_t (\varphi)=\varphi^0_tB^h_t=\FCshk_t$ for all $t\in [0,\Tdelta]$ and thus it is clear that $\varphi^0_t =(B^h_t)^{-1}\FCshk_t$ for all $t\in [0,\Tdelta]$. To compute the hedge ratio $\varphi^1_t$, it suffices to identify the martingale part
in the dynamics of the price process $\FCshk$. We obtain, for all $t\in [0,\Tdelta]$,
\begin{align*}
d\FCshk_t &\stackrel{mart}{=} N\big(h_+(t)\big)\,d\Gamma_t  - N\big(h_{-}(t)\big)\,d\Lambda_t
+\Gamma_t\,dN\big(h_+(t)\big) -\Lambda_t\,dN\big(h_{-}(t)\big)
\end{align*}
where the equality $\stackrel{mart}{=}$ means that the processes have identical continuous martingale part.
It is easy to check that
\begin{align*}
d\Gamma_t \stackrel{mart}{=} -  \I_{[0,U]}(t) \sigma n(t,U) \Gamma_t \,dW_t^{\Qf}, \quad
d\Lambda_t \stackrel{mart}{=} - \sigma n(t,T) \Lambda_t \,dW_t^{\Qf}.
\end{align*}
Furthermore,
\begin{align*}
 dN\big(h_{\pm}(t)\big)\stackrel{mart}{=}n\big(h_{\pm}(t)\big)\,\frac{d(1+\delta \Rsf_t)}{v_Y(t,\Udelta,\Tdelta)(1+\delta \Rsf_t)}
\end{align*}
and thus using \eqref{eq3.13} and \eqref{eq3.14} we obtain, for all $t\in [0,\Udelta]$,
\begin{align*}
dN\big(h_{\pm}(t)\big)\stackrel{mart}{=} n\big(h_{\pm}(t)\big)(v_Y(t,\Udelta,\Tdelta))^{-1} n(t,\Udelta,\Tdelta)\sigma \, dW^{\Qf}_t
\end{align*}
and,  for all $t\in [\Udelta,\Tdelta]$,
\begin{align*}
dN\big(h_{\pm}(t)\big)\stackrel{mart}{=} n\big(h_{\pm}(t)\big)(v_Y(t,\Udelta,\Tdelta))^{-1} n(t,\Tdelta)\sigma\, dW^{\Qf}_t.
\end{align*}
The asserted equalities for $\varphi^1$ can now be obtained by straightforward computations.
\endproof

\subsection{Pricing and Hedging of SOFR Swaptions}   \label{sec3.8}

In contrast to the case of a caplet, the arbitrage-free pricing and hedging of a SOFR swaption (that is, an option written on a multi-period SOFR swap) is more computationally challenging but still feasible within the present setup. In fact,
in our analysis of a SOFR swaption we may use some arguments from the classical case of bond options and swaptions referencing LIBOR within the framework of a single factor model (see, e.g., Jamshidian \cite{J1989} for bond options and H\"ubner \cite{H1996} for LIBOR swaptions).

\begin{definition}\label{def4.1} {\rm
The payoff of a {\it SOFR swaption} with expiration date $T_0$ and strike $\kappa>0$ equals $(\Swapn_{T_0}(T_0,n))^+$.}
\end{definition}

We consider the case of an uncollateralized swaption but it is rather clear that the case of a swaption with a proportional collateralization can be dealt with using the method from Section \ref{sec3.6}.
The arbitrage-free price of an uncollateralized SOFR swaption is given by the following expression
\begin{align*}
\PSshk_t(T_0,n)=\Bhh_t \,\EQf\Big((\Bhh_{T_0})^{-1}\big(\Swapn_{T_0}(T_0,n)\big)^+\,\big|\,\cF_t\Big)
\end{align*}
where (see Proposition \ref{pro3.6})
\begin{align*}
\Swapn_{T_0}(T_0,n)=\sum_{j=1}^n\Big(A^{s,h}(T_{j-1},T_j)\Bhh(T_0,T_{j-1})-(1+\delta_j \kappa)\Bhh(T_0,T_j)\Big).
\end{align*}
In all results pertaining to the pricing and hedging of SOFR swaptions, we assume that $\alpha^h (t)\geq \alpha^s (t)$ for every $t \in [\Tdelta_0,\Tdelta_n]$ and thus $0<A^{s,h}(T_{j-1},T_j)\leq 1$ for every $j=1,2,\dots,n$ or, equivalently,  $0 \leq a_{j-1}:=1-A^{s,h}(T_{j-1},T_j)<1$. By convention, we also set $a_n=1$.

Let us first examine the exercise event $\{\Swapn_{T_0}(T_0,n)>0\}$ of the SOFR swaption. The decision to either exercise or abandon a swaption by its holder relies on the arbitrage-free price of the underlying SOFR swap at the swaption's maturity date $T_0$ but it can also be expressed in terms of $x_{T_0}$, specifically, $\{\Swapn_{T_0}(T_0,n)>0\}=\{x_{T_0}>x^*\}$ where $x^*$ solves equation \eqref{eqxx}.

\begin{lemma}
Assume that $\alpha^h (t)\geq \alpha^s (t)$ for every $t \in [\Tdelta_0,\Tdelta_n]$. Then a SOFR swaption is exercised at expiration date $T_0$ if and only if $x_{T_0}>x^*$ where a real number $x^*=x^*(T_0,T_1,\dots,T_n)$
is a unique solution to the equation
\begin{align}  \label{eqxx}
\sum_{j=1}^n c_j A^h(T_0,\Tdelta_j) e^{m(T_0,\Tdelta_j)-n(T_0,\Tdelta_j)x}=c_0
\end{align}
where $c_0:= 1-a_0, c_j:=a_j+\delta_j\kappa$ for $j=1,2,\dots, n-1$ and $c_n:=a_n+ \delta_n\kappa = 1 +\delta_n\kappa$.
\end{lemma}

\proof
The swaption's pricing formula can be rewritten as, for all $t\in [0,T_0]$,
\begin{align*}
\PSshk_t(T_0,n)=  \Bhh_t \, \EQf \bigg[ (\Bhh_{T_0})^{-1}
\bigg(\sum_{j=1}^n\Big( (1-a_{j-1}) \Bhh(T_0,T_{j-1})-(1 +\delta_j \kappa)\Bhh(T_0,T_j) \Big)\bigg)^+ \,\Big|\, \cF_t\bigg]
\end{align*}
or, equivalently,
\begin{align} \label{eqq1}
\PSshk_t(T_0,n)=\Bhh_t\,\EQf\bigg[(\Bhh_{T_0})^{-1}\bigg(c_0-\sum^n_{j=1}c_j\Bhh(T_0,T_j)\bigg)^+ \,\Big|\, \cF_t\bigg]
\end{align}
where we denote $c_0:= 1-a_0, c_j:=a_j+\delta_j\kappa$ for $j=1,2,\dots, n-1$ and $c_n:=a_n+ \delta_n\kappa=1 +\delta_n\kappa$.
Notice that $c_j$ is a positive constant for every $j=0,1,\dots,n$ since $\kappa $ is assumed to be positive.
Recall from Lemma \ref{lem3.2} that for every $j=1,2,\dots, n$, the process $\Bhh(t,\Tdelta_j)$ equals, for every $t\in[0,T_0]$,
\begin{align} \label{eqq2}
\Bhh(T_0,\Tdelta_j)=A^h(T_0,\Tdelta_j) e^{m(T_0,\Tdelta_j)-n(T_0,\Tdelta_j)x_{T_0}}=: g_j(x_{T_0}).
\end{align}
Consequently, the function $g: \mathbb{R} \to \mathbb{R}$ given by $g(x) =\sum^n_{j=1}c_j g_j(x)$ is strictly
decreasing and has the limits $\lim_{x\to -\infty} g(x)= \infty $ and  $\lim_{x\to \infty} g(x)=0$.
Therefore, for any $c_0 \in (0,1)$ there exists a unique solution $x^*$ to the equation $g(x)=c_0$. Furthermore, $g(x)<c_0$ for every $x>x^*$, which implies that the SOFR swaption expires in the money if and only if $x_{T_0}>x^*$.
\endproof

\begin{remark} {\rm \label{rem4.1xv}
Recall from \eqref{nUT} that
\begin{align*}
n(T_0,\Tdelta_{j-1},\Tdelta_j)x_{T_0}=\ln\big[A^s(\Tdelta_{j-1},\Tdelta_j)\big(1+\delta_j \Rsf_{T_0}(\Tdelta_{j-1},\Tdelta_j)\big)\big]-\Theta (T_0,\Tdelta_{j-1},\Tdelta_j)
\end{align*}
where $n(T_0,\Tdelta_{j-1},\Tdelta_j)$ is strictly positive for every $j=1,2,\dots,n$.
Hence the inequality $x_{T_0}>x^*$ holds whenever $\Rsf_{T_0}(\Tdelta_{j-1},\Tdelta_j)>r^*_j$ where $r^*_j$ satisfies
\begin{align*}
1+\delta_j r^*_j = (A^s(\Tdelta_{j-1},\Tdelta_j))^{-1}e^{n(T_0,\Tdelta_{j-1},\Tdelta_j)x^* +\Theta (T_0,\Tdelta_{j-1},\Tdelta_j)}
\end{align*}
or, equivalently, the inequality $f^{s}_{T_0}(\Tdelta_{j-1},\Tdelta_j)< 1-r^*_j$ is satisfied
by the SOFR futures price $f^{s}_{T_0}(\Tdelta_{j-1},\Tdelta_j)=1-\Rsf_{T_0}(\Tdelta_{j-1},\Tdelta_j)$.
} \end{remark}

We are in a position to establish the pricing formula for a SOFR swaption.

\begin{proposition}\label{pro3.12}
Assume that $\alpha^h (t)\geq \alpha^s (t)$ for every $t \in [\Tdelta_0,\Tdelta_n]$. Then the arbitrage-free price
of a SOFR swaption with the expiration date $T_0$ and strike $\kappa>0$ equals, for every $t\in [0,\Tdelta_0]$,
\begin{align*}
\PSshk_t(T_0,n)&= e^{-n(t,T_0)x_t+\eta(t,T_0)}c_0 A^h(t,T_0)N(-\wh{x}_0) \\
&\quad - e^{-n(t,T_0)x_t+\eta(t,T_0)} \sum^n_{j=1}c_jA^h(t,\Tdelta_j)
   e^{m(T_0,\Tdelta_j)-n(T_0,\Tdelta_j)\mu(x_t)+\frac{1}{2}n^2(T_0,\Tdelta_j)v_0^2}N(-\wh{x}_j)
\end{align*}
where $c_0:= 1-a_0, c_j:=a_j+\delta_j\kappa$ for $j=1,2,\dots, n-1$ and $c_n:=a_n+\delta_n\kappa = 1 +\delta_n\kappa$.
Furthermore,
\begin{align*}
\eta(t,T_0)&:=-\int_t^{T_0}a n(u,T_0)\,du+\frac{1}{2}\int_t^{T_0}\sigma^2 n^2(u,T_0)\,du, \\
\mu(x_t)&:= 
e^{-b(T_0-t)}x_t+\int_t^{T_0}e^{-b(T_0-u)}\wh{a}(u)\,du,\\
v^2_0&:=\int_t^{T_0}e^{-2b(T_0-u)}\sigma^2\,du,
\end{align*}
where $\wh{a}(t):=a-\sigma^2 n(t,T_0)$. Finally, we denote $\wh{x}_j:=\wh{x}+n(T_0,\Tdelta_j)v_0$ for every $j=0,1,\dots,n$ where $\wh{x}:=(x^*-\mu(x_t))v_0^{-1}$.
\end{proposition}

\proof
Recall that
\begin{align*}
\Bhh_t(\Bhh_{T_0})^{-1}=e^{-\int_t^{T_0}r^h_u\,du}=A^h(t,T_0)e^{-\int_t^{T_0}x_u\,du}=A^h(t,T_0)e^{-\xxx_{t,T_0}}
\end{align*}
where (see the proof of Lemma \ref{lem3.1})
\begin{align*}
\xxx_{t,T_0}:=\int_t^{T_0}x_u\,du=n(t,T_0)x_t+\int_t^{T_0}a n(u,T_0)\,du+\int_t^{T_0}\sigma n(u,T_0)\,dW^{\Qf}_u.
\end{align*}
We define the probability measure $\Qhf$ equivalent to $\Qf$ on $(\Omega,\cF_{T_0})$
by the Radon-Nikodym density $d\Qhf/d\Qf=M_{T_0}$ where the continuous, square-integrable martingale $M_t,\, t \in [0,T_0]$ is given by
\begin{align*}
M_t:=e^{-\int_0^t\sigma n(u,T_0)\,dW^{\Qf}_u-\frac{1}{2}\int_0^t\sigma^2 n^2(u,T_0)\,du}
\end{align*}
so that the process $W^{\Qhf}_t=W^{\Qf}_t+\int_0^t\sigma n(u,T_0)\,du,\,t\in [0,T_0]$ is a Brownian motion under
$\Qhf$. Then equations \eqref{eqq1} and \eqref{eqq2} yield
\begin{align*}
\PSshk_t(T_0,n)=C_t\,\EQhf\bigg[\bigg(c_0-\sum^n_{j=1}c_j\Bhh(T_0,T_j)\bigg)^+\,\Big|\,\cF_t\bigg]
 = C_t\,\EQhf\bigg[\bigg(c_0-\sum^n_{j=1}c_j g_j(x_{T_0})\bigg)^+\,\Big|\,\cF_t\bigg]
\end{align*}
where the $\mathcal{F}_t$-measurable random variable $C_t$ is given by
\begin{align*}
C_t:=A^h(t,T_0)e^{-n(t,T_0)x_t-\int_t^{T_0}a n(u,T_0)\,du+\frac{1}{2}\int_t^{T_0}\sigma^2 n^2(u,T_0)\,du}
    =A^h(t,T_0)e^{-n(t,T_0)x_t+\eta (t,T_0)}.
\end{align*}

The factor process $x$ is governed under $\Qhf$ by the extended Vasicek's equation, for all $t\in [0,T_0]$,
\begin{align} \label{eq3.10c}
dx_t=(\wh{a}(t)-b x_t)\,dt+\sigma\,dW_t^{\Qhf}
\end{align}
where $\wh{a}(t):=a-\sigma^2 n(t,T_0)$. By solving equation \eqref{eq3.10c}, we obtain
\begin{align*}
x_{T_0}=e^{-b(T_0-t)}x_t+\int_t^{T_0}e^{-b(T_0-u)}\wh{a}(u)\,du+\int_t^{T_0}e^{-b(T_0-u)}\sigma\,dW_u^{\Qhf},
\end{align*}
which shows that the $\cF_t$-conditional distribution of the random variable $x_{T_0}$ under $\Qhf$ is the Gaussian distribution $N(\mu(x_t),v_0^2)$ where $\mu(x_t)$ and $v_0^2$ are given in the statement of the proposition.
Consequently,
\begin{align*}
&\PSshk_t(T_0,n)=C_t\int_{x^*}^{\infty}\bigg(c_0-\sum^n_{j=1}c_jA^h(T_0,\Tdelta_j)
    e^{m(T_0,\Tdelta_j)-n(T_0,\Tdelta_j)y} \bigg)\frac{1}{\sqrt{2\pi}v_0} e^{-\frac{(y-\mu(x_t))^2}{2v_0^2}}\,dy.
\end{align*}
Upon setting $z=(y-\mu(x_t))v_0^{-1}$ and $\wh{x}=(x^*-\mu(x_t))v_0^{-1}$, we obtain ($n(z)$ is the standard Gaussian p.d.f.)
\begin{align*}
&\PSshk_t(T_0,n)=C_t\int_{\wh{x}}^{\infty}\bigg(c_0-\sum^n_{j=1}c_jA^h(T_0,\Tdelta_j)
    e^{m(T_0,\Tdelta_j)-n(T_0,\Tdelta_j)(v_0 z+\mu(x_t))}\bigg) n(z)\,dz \\
&=A^h(t,T_0)e^{-n(t,T_0)x_t+\eta (t,T_0)}\int_{\wh{x}}^{\infty}\bigg(c_0-\sum^n_{j=1}c_jA^h(T_0,\Tdelta_j)
    e^{m(T_0,\Tdelta_j)-n(T_0,\Tdelta_j)(v_0 z+\mu(x_t))}\bigg) n(z)\,dz \\
&=e^{-n(t,T_0)x_t+\eta(t,T_0)} c_0 A^h(t,T_0) \int_{\wh{x}}^{\infty}n(z)\,dz\\
&\quad - e^{-n(t,T_0)x_t+\eta(t,T_0)}\sum^n_{j=1}c_jA^h(t,\Tdelta_j)e^{m(T_0,\Tdelta_j)-n(T_0,\Tdelta_j)\mu(x_t)}
   \int_{\wh{x}}^{\infty}e^{-n(T_0,\Tdelta_j)v_0 z} n(z)\,dz \\
&= e^{-n(t,T_0)x_t+\eta(t,T_0)}c_0 A^h(t,T_0)N(-\wh{x}_0) \\
&\quad - e^{-n(t,T_0)x_t+\eta(t,T_0)} \sum^n_{j=1}c_jA^h(t,\Tdelta_j)
   e^{m(T_0,\Tdelta_j)-n(T_0,\Tdelta_j)\mu(x_t)+\frac{1}{2}n^2(T_0,\Tdelta_j)v_0^2}N(-\wh{x}_j)
\end{align*}
where we denote $\wh{x}_j:=\wh{x}+n(T_0,\Tdelta_j)v_0$ and the last equality holds since
\begin{align*}
\int_{\wh{x}}^{\infty}e^{-n(T_0,\Tdelta_j)v_0 z}n(z)\,dz
 = e^{\frac{1}{2}n^2(T_0,\Tdelta_j)v_0^2}\int_{\wh{x}+n(T_0,\Tdelta_j)v_0}^{\infty} n(z)\,dz
 = e^{\frac{1}{2}n^2(T_0,\Tdelta_j)v_0^2}N(-\wh{x}_j).
\end{align*}
Furthermore, we have that $m(T_0,T_0)=n(T_0,T_0)=0$ and thus the proof of the proposition is completed.
\endproof

It follows from Proposition \ref{pro3.12} that the arbitrage-free price of a SOFR swaption has the following representation
$\PSshk_t(T_0,n)=\PS^0_t(T_0,n)-\sum_{j=1}^n \PSj_t(T_0,n)$ where
\begin{align}  \label{4.10}
\PSj_t(T_0,n)=k_j(t)e^{l_j(t) x_t}N\big(\mu_j(x_t)\big)
\end{align}
and we denote
\begin{align*}
k_j(t)&:= c_j A^h(t,\Tdelta_j)e^{\eta(t,T_0)+m(T_0,\Tdelta_j)
-n(T_0,\Tdelta_j)\int_t^{T_0}e^{-b(T_0-u)}\wh{a}(u)\,du+\frac{1}{2}n^2(T_0,\Tdelta_j)v_0^2},\\
l_j(t)&:=-n(t,T_0)-n(T_0,\Tdelta_j)e^{-b(T_0-t)}, \\
\mu_j(x_t)&:= (\mu(x_t)-x^*)v_0^{-1}-n(T_0,\Tdelta_j)v_0,
\end{align*}
where as before $\wh{a}(t):=a-\sigma^2 n(t,T_0)$.
It is natural to postulate that the payoffs $X_0:=\PS^0_{T_0}(T_0,n)$ and $X_1 =-\PS^1_{T_0}(T_0,n)$
should be hedged using the SOFR futures contracts referencing the accrual period $[T_0,T_1]$
and the payoff $X_j:=-\PSj_{T_0}(T_0,n)$ for each $j=2,3,\dots,n$ should be hedged using the SOFR futures contracts referencing the accrual period $[T_{j-1},T_j]$. Notice that $\PSshk_{T_0}(T_0,n)=\sum_{j=0}^n X_j$ and thus it suffices
to fix an arbitrary $j \in \{0,1,2,\dots ,n\}$ and examine the replicating strategy for the payoff $X_j$.

\begin{proposition}\label{pro3.1b}
The payoff $X_0=\PS^0_{T_0}(T_0,n)$ can be replicated by a futures trading strategy $(\varphi^{0,0},\varphi^0)$ where
the SOFR futures price $f^{s,0}=f^s(T_0,T_1)$ references $[T_0,T_1]$ and for all $t \in [0,\Tdelta_0]$ we have $\varphi^{0,0}_t \Bhh_t=\PS^0_t(T_0,n)$ and
\begin{align*}
\varphi^0_t=-\frac{k_0(t) e^{l_0(t)x_t} \big[ l_0(t) N\big(\mu_0(x_t)\big)
+e^{-b(T_0-t)}v_0^{-1}n\big(\mu_0(x_t)\big)\big]}{\delta_1^{-1}\big(1+\delta_1\Rsf_t(\Tdelta_0,\Tdelta_1)\big)
n(t,\Tdelta_0,\Tdelta_1)}.
\end{align*}
For every $j=1,2,\dots,n$, the payoff $X_j=-\PSj_{T_0}(T_0,n)$ can be replicated by a futures trading strategy $(\varphi^{0,j},\varphi^j)$ where the SOFR futures price $f^{s,j}=f^s(T_{j-1},T_j)$ references $[T_{j-1},T_j]$ and for all $t \in [0,\Tdelta_0]$ we have $\varphi^{0,j}_t \Bhh_t=-\PSj_t(T_0,n)$ and
\begin{align*}
\varphi^j_t=\frac{k_j(t) e^{l_j(t)x_t} \big[ l_j(t) N\big(\mu_j(x_t)\big)
+e^{-b(T_0-t)}v_0^{-1}n\big(\mu_j(x_t)\big)\big]}{\delta_j^{-1}\big(1+\delta_j\Rsf_t(\Tdelta_{j-1},\Tdelta_j)\big)
n(t,\Tdelta_{j-1},\Tdelta_j)}.
\end{align*}
Then the futures trading strategy $\big(\sum_{j=0}^n \varphi^{0,j},\varphi^0,\varphi^1,\dots ,\varphi^n\big)$ where the SOFR futures price $f^{s,0}=f^s(T_0,T_1)$ references the period $[T_0,T_1]$ and, for every $j=1,2,\dots,n$, the SOFR futures price $f^{s,j}=f^s(T_{j-1},T_j)$ references the period $[T_{j-1},T_j]$ replicates the payoff $\PSshk_{T_0}(T_0,n)$ of the SOFR swaption.
\end{proposition}

\proof
On the one hand, from  \eqref{4.10} we obtain, for every $j=0,1,\dots,n$,
\begin{align} \label{4.11}
d\PSj_t(T_0,n)&\stackrel{mart}{=}k_j(t) e^{l_j(t)x_t} \big[ l_j(t) N\big(\mu_j(x_t)\big)
+e^{-b(T_0-t)}v_0^{-1}n\big(\mu_j(x_t)\big)\big]\sigma \,dW_t^{\Qf}.
\end{align}
On the other hand, it is known from Lemma \ref{lem3.1} that the dynamics of the SOFR futures price $f^{s,j}_t:=f^{s}_t(\Tdelta_{j-1},\Tdelta_j)$ for $t \in [0,T_{j-1}]$ are (see \eqref{fut3.22})
\begin{align*}
df^{s,j}_t=-\delta_j^{-1}\big(1+\delta_j\Rsf_t(\Tdelta_{j-1},\Tdelta_j)\big)n(t,\Tdelta_{j-1},\Tdelta_j)\sigma\,dW^{\Qf}_t
\end{align*}
and thus the wealth process of a self-financing futures strategy $(\varphi^{0,j},\varphi^j)$ satisfies under $\Qf$ (see \eqref{fut3.22x})
\begin{align} \label{4.12}
dV_t(\varphi^j)=\varphi^{0,j}_t\,d\Bhh_t-\varphi^j_t \delta_j^{-1}\big(1+\delta_j\Rsf_t(\Tdelta_{j-1},\Tdelta_j)\big)n(t,\Tdelta_{j-1},\Tdelta_j)\sigma\,dW^{\Qf}_t.
\end{align}
Consequently, it is easy to identify the process $\varphi^j$ by comparing the diffusion terms in the dynamics
of $\PSj(T_0,n)$ and $V(\varphi^j)$ as given by \eqref{4.11} and \eqref{4.12}, respectively. Then the
equalities  $\varphi^{0,0}_t \Bhh_t=\PS^0_t(T_0,n)$ and $\varphi^{0,j}_t \Bhh_t=-\PSj_t(T_0,n)$ for $j=1,2,\dots,n$
are satisfied for every $t \in [0,\Tdelta_0]$ and thus we conclude
that $\sum_{j=0}^n \varphi^{0,j} \Bhh_t=\PSshk_t(T_0,n)$ for every $t \in [0,\Tdelta_0]$, which means that
the futures trading strategy from thee statement of the proposition replicates the SOFR swaption.
\endproof

\begin{remark}
{\rm It is clear that it is also possible to replicate the payoff $X_0=\PS^0_{T_0}(T_0,n)$ using the SOFR futures contracts referencing the accrual period $[T_{-1},T_0]$ where $T_{-1}=T_0-\delta_0$.
Then the hedge ratio $\varphi^0_t$ appearing in Proposition \ref{pro3.1b} would need to be modified accordingly
using equations from Lemma \ref{lem3.1}, which means that it would have a slightly different representation for $t\in [0,T_{-1}]$ and for $t\in [T_{-1},T_0]$. However, this would be a dubious choice of a hedging instrument for the SOFR swaption since the accrual period $[T_{-1},T_0]$ is not relevant for its payoff. Even more importantly, it would be good
to extend the model so that the SOFR futures prices referencing different accrual periods would not be perfectly correlated.
}
\end{remark}

\section{Numerical Results} \label{sec5}

We conclude this study by presenting a numerical analysis of various SOFR swap classes within the framework of the Vasicek model. This section is structured as follows: first, we introduce the model parameters using and examine comparative statics. Next, we examine the impact of varying the basis spread $\kappa$, on the arbitrage-free pricing of SOFR swaps, SOFR caps, and SOFR swaptions. Finally, we evaluate the risk exposure of SOFR swaps by analyzing profit and loss (P\&L) profiles for unhedged and hedged positions under varying hedging frequencies, providing insights into the effectiveness of different hedging strategies.

As a representative example, we consider a standard 3-year multi-period SOFR swap with the notional principal of 10 million. This swap is a fixed-for-floating arrangement with the floating rate determined by the backward-looking compound SOFR average. As in Definition \ref{def3.3}, the structure of the SOFR swap is specified as follows: $T_0$ denotes the inception date and $T_n$ is the maturity date with $n = 6$. Hence the tenor structure is given by $0 < T_1 < \cdots < T_6$ where $T_{j} = 0.5j$ for $j = 1,2,\dots, 6$ and $\delta_j := T_{j} - T_{j-1} = 0.5$ for all $j$. At each payment date $T_{j}$ for $j = 1,2, \dots, 6$ the net cash flow for a SOFR swap with the notional principal $P = 10$ million is given by
\begin{equation*}
\FS^{s,\kappa}_{T_j}(T_{j-1}, T_j) = \big( R^s(T_{j-1}, T_j)-\kappa \big)\delta_j\Noti
\end{equation*}
where $[T_{j-1}, T_j]$ is the $j$th accrual period. For simplicity, we first assume that $T_0 = 0$, which means
that the contract is initiated at time 0, as opposed to a forward swap starting at some future date $T_0 > 0$.

We focus on a multi-period uncollateralized SOFR swap, as discussed in Section \ref{sec3.4}, with $\beta = 0$. Additionally, we assume that $\alpha^s = 0$, which means that the SOFR plays the role of the base rate. For the sake of comparison of theoretical pricing formulae with the Monte Carlo simulations, we use the following annualized parameters from Assumption \ref{ass3.1}:
\[
a = 0.1, \quad b = 5, \quad \sigma = 1\%, \quad x_0 = 2\%, \quad \alpha^h = 1\%.
\]
The correctness of theoretical pricing formulae is first validated using the Monte Carlo method. The numerical prices computed for SOFR swaps, SOFR futures options, SOFR caps, and SOFR swaptions exhibit an almost perfect agreement with the theoretical prices, with only negligible simulation errors, thus yielding a solid confirmation of the validity and reliability of our pricing framework.

Our first objective is to illustrate the pricing outcomes discussed in Section \ref{sec3.3}. To facilitate an analysis, we rewrite the Vasicek dynamics from Assumption \ref{ass3.1} as
\[
dx_t = b (\theta - x_t)\, dt + \sigma \, dW_t^{\Qf}
\]
where $\theta := a/b$ represents the long term mean level of the factor process and $b$ characterizes the recovery speed.
We aim to analyze the effect of varying model parameters, particularly the recovery speed $b$ and the long-term mean level of the interest rate $\theta$, on the arbitrage-free pricing of a multi-period SOFR swap derived in Proposition \ref{pro3.6}.

In Table \ref{Table: parameters study multiperiod}, we report the arbitrage-free prices, denoted by $\Swapn_0(T_0, n)$, and the corresponding fair basis spread, $\kappa$, at the inception date $T_0 = 0$ for a standard SOFR swap with the notional principal of 10 million. Results in Table \ref{Table: parameters study multiperiod} highlight the intricate relationship between model parameters and prices of SOFR swaps. The long term mean level, $\theta$, emerges as a dominant factor influencing both the fair price and thus basis spread, as demonstrated in cases 1, 2, and 5. This effect is nearly linear, reflecting the dominant role of $\theta$ in determining the overall trajectory of the interest rate process and, consequently, the swap’s valuation.

\begin{table}[htbp]
\centering
\caption{Impact of the model parameters on the arbitrage-free price and fair basis spread}
\label{Table: parameters study multiperiod}
\begin{tabular}{cccccc}
\toprule
No. & $b$ & $\theta$ & $\sigma$ & $\Swapn_0 \times 10^7$ & Basis spread $\kappa$\\
\midrule
1 & $1$ & $2\%$ & $1\%$ & $571,620$ & $200.74$ bps \\
2 & $5$ & $2\%$ & $1\%$ & $572,307$ & $200.99$ bps \\
3 & $10$ & $2\%$ & $1\%$ & $572,343$ & $201.00$ bps \\
4 & $5$ & $1\%$ & $1\%$ & $309,689$ & $107.10$ bps \\
5 & $5$ & $5\%$ & $1\%$ & $1,317,645$ & $484.33$ bps \\
6 & $5$ & $2\%$ & $5\%$ & $571,108$ & $200.55$ bps \\
7 & $5$ & $2\%$ & $10\%$ & $567,363$ & $199.19$ bps \\
8 & $1$ & $2\%$ & $10\%$ & $498,428$ & $174.45$ bps \\
\bottomrule
\end{tabular}
\end{table}

By contrast, the recovery speed, $b$, and the volatility, $\sigma$, exhibit limited influence when operating within reasonable ranges. This is evident from cases 1, 3, 6, and 7, where fixing one of these parameters while varying the other yields only marginal changes in the pricing outcomes. This observation aligns with theoretical expectations since, for a swap contract with a long term horizon, the dynamics are primarily driven by the mean-reverting property of the Vasicek process, rendering $b$ and $\sigma$ less impactful under normal conditions.

\begin{figure}[htbp]
\centering
\includegraphics[width=0.7\linewidth]{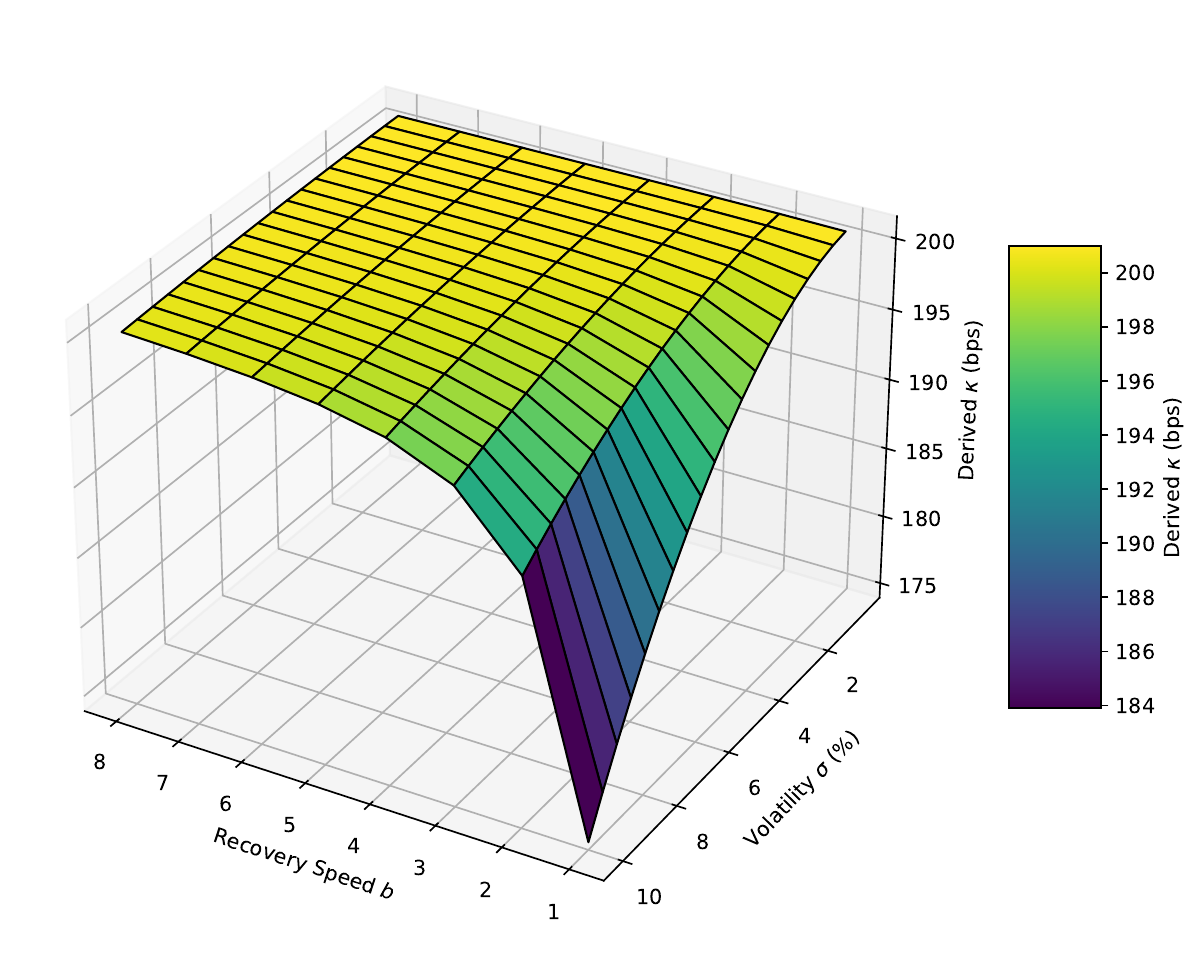}
\caption{Impact of the speed of recovery and volatility on the fair basis spread}
\label{fig: kappa_sigma_b}
\end{figure}

However, extreme values of $\sigma$ combined with low values of recovery speed $b$ can lead to a significant deviation from the standard case, as illustrated in case 8 and visualized in Figure \ref{fig: kappa_sigma_b}. Under such circumstances, the heightened volatility and slow mean-reversion create substantial uncertainty in the interest rate’s path, causing the market to adjust its valuation of the fair price and basis spread accordingly. This behavior underscores the importance of considering parameter extremes in practical applications, where market conditions may deviate from typical assumptions.

To investigate the comparative pricing of swaps, caps, and swaptions referencing the SOFR under varying values of $\kappa$, we begin by modifying the contract structure to ensure meaningful results. Specifically, we shift the inception date forward by six months, setting $T_0 = 0.5$ and defining the maturity dates as $T_j = 0.5(j + 1)$ for $j = 1,2, \dots, 6$, while keeping all other parameters unchanged. This adjustment transforms the contracts into forward-starting instruments, where the inception date $T_0$ lies in the future ($T_0 > 0$).

The parameter $\kappa$ is initially set to a default value of $200$ bps. To investigate its effect, we perturb $\kappa$ above and below this baseline and hence introduce the variability into the pricing outcomes. Our earlier analysis highlighted the role of $\theta$ as a key driver in the pricing framework so we fix $\theta$ at $2\%$ in the present analysis and we
examine the influence of $\kappa$ on prices of SOFR derivatives.

\begin{table}[htbp]
\centering
\caption{Impact of $\kappa$ on arbitrage-free prices for SOFR derivatives}
\label{Table: parameters study three prices}
\begin{tabular}{ccccc}
\toprule
No. & Basis spread $\kappa$ & $\Swapn_0 \times 10^7$ & $\PSshk_0 \times 10^7$ & $\FCshk_0 \times 10^7$ \\
\midrule
1 & $150$ bps & $143,030$ & $143,030$ & $143,307$ \\
2 & $190$ bps & $30,602$ & $33,348$ & $43,619$ \\
3 & $200$ bps & $2,774$ & $14,062$ & $26,894$ \\
4 & $210$ bps & $-25,277$ & $3,812$ & $14,820$ \\
5 & $250$ bps & $-137,482$ & $0$ & $358$ \\
\bottomrule
\end{tabular}
\end{table}

As anticipated, the prices of SOFR swaps, denoted by $\Swapn_0$, exhibit approximate symmetry around the benchmark value of $\kappa = 200$ bps. In contrast, the prices for SOFR caps and swaptions are monotonically decreasing as $\kappa$ increases, which is consistent with theoretical properties. Furthermore, we observe a dominance relationship among the three instruments, specifically, we have that $\Swapn_0 \leq \PSshk_0 \leq \FCshk_0$. In fact, this ordering arises naturally from the structural properties of these contracts and is preserved under the expectation operator.

In the last step of our numerical study, we investigate the risk exposure associated with SOFR swaps through the simulation of P\&L profiles for both unhedged and hedged positions, under varying hedging frequencies. This analysis builds on discretization of hedging strategies discussed in Section \ref{sec3.5}, which rely on the SOFR futures dynamics developed in Section \ref{sec3.2}. Specifically, we simulate the futures price dynamics and subsequently derive the wealth process for a discretized replicating strategy, thereby quantifying the effectiveness of hedging in reducing risk exposure.

To assess the impact of hedging frequency on the P\&L distribution, we consider hedging strategies with weekly, monthly, and biannual rebalancement of hedging portfolio. By analyzing the P\&L quantiles, we aim to provide a comprehensive view of the asset's riskiness and we use the 25\% and 75\% quantiles as indicative measures of risk exposure. These quantiles capture the variability in P\&L outcomes, reflecting the potential financial impact of market fluctuations under different hedging regimes.

\begin{figure}[tbhp]
    \centering
    \subfloat[Weekly rehedged P\&L plot]{
        \label{fig:weekly_hedged_PnL}
        \includegraphics[width=0.45\textwidth]{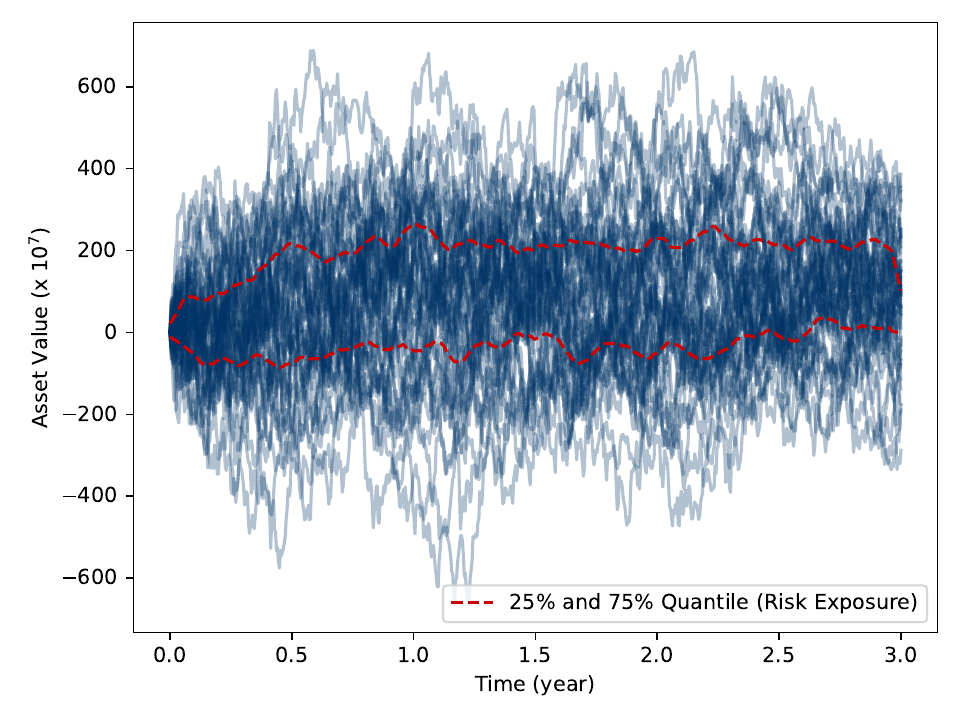}
    }
    \hfill
    \subfloat[Monthly rehedged P\&L plot]{
        \label{fig:month_hedged_PnL}
        \includegraphics[width=0.45\textwidth]{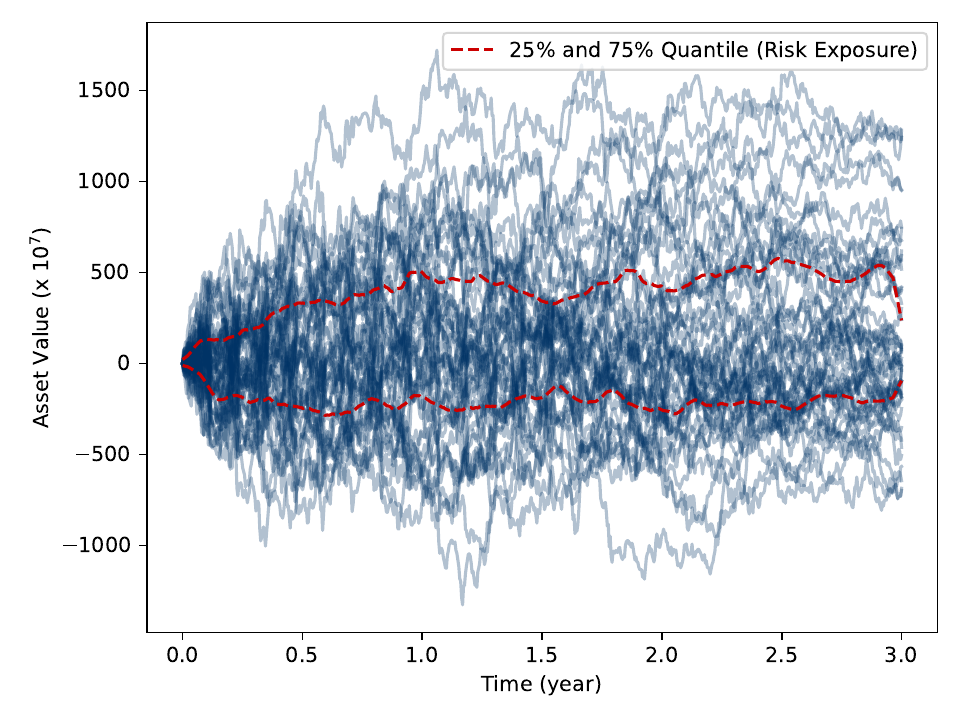}
 }

    \medskip

    \subfloat[Biannually rehedged P\&L plot]{
        \label{fig:biannual_hedged_PnL}
        \includegraphics[width=0.45\textwidth]{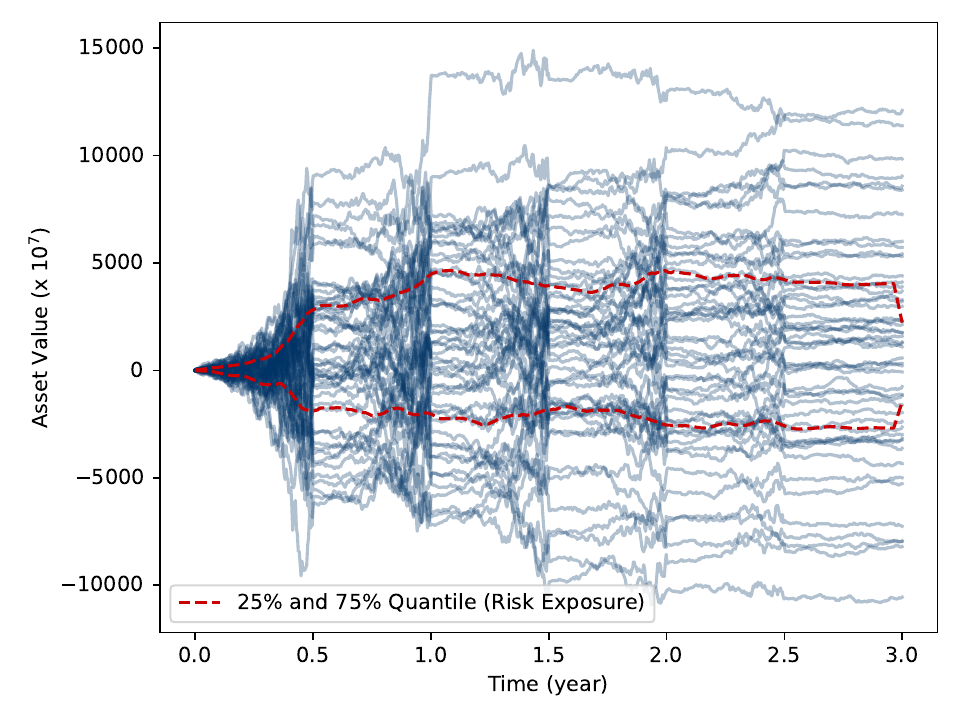}
    }
     \hfill
      \subfloat[Unhedged P\&L plot]{
        \label{fig:unhedged_PnL}
        \includegraphics[width=0.45\textwidth]{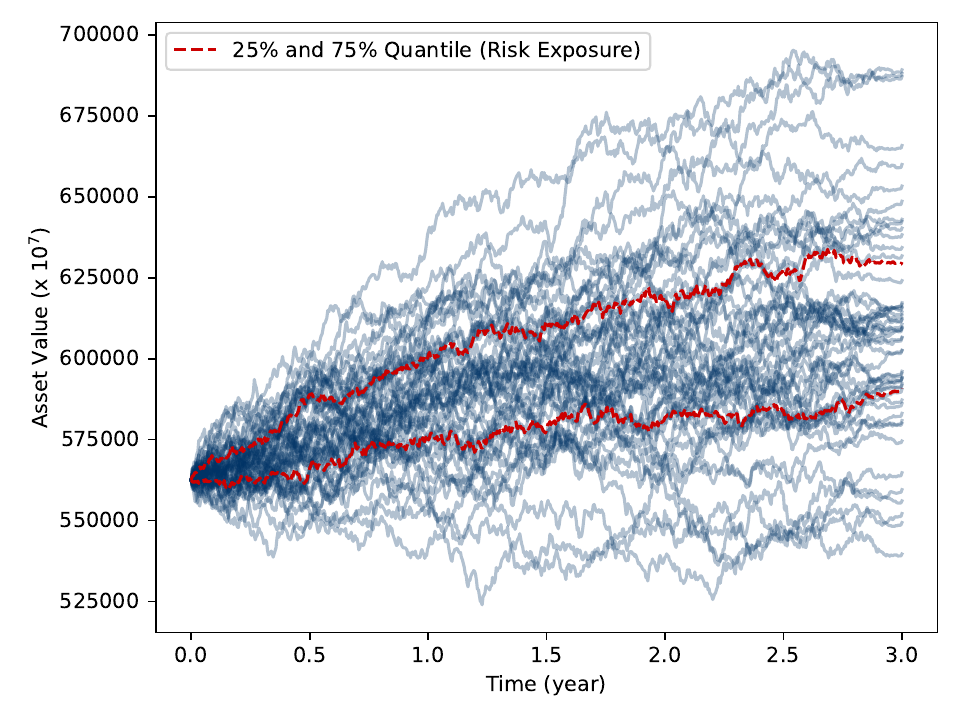}
    }
    \caption{Comparison of P\&L plots for different rehedging frequencies}
    \label{fig:different_hedge}
\end{figure}

Across all hedging strategies, the distance between the 25\% and 75\% quantiles tends to increase over time, which indicates
a progressive growth in risk exposure. In the hedged scenarios, this expansion in risk exposure is sharply reduced immediately following each hedging action. Between these rehedging points, however, the risk exposure gradually increases, reflecting the accumulation of unhedged risks. This behavior is particularly pronounced in Figure \ref{fig:different_hedge}(c), where the semiannual hedging strategy is characterised by the most significant fluctuations between rebalancing intervals.

As expected, the effectiveness of the hedging strategy is positively correlated with the frequency of rebalancing. Moving from the weekly to monthly, and then to biannual rehedging, the risk exposure systematically increases. This trend underscores the importance of frequent rebalancing in minimizing the risk exposure. Collectively, these findings validate the practical effectiveness of the developed hedging strategies and hence support the theoretical results presented in Section~\ref{sec3}.

\newpage

\section{Concluding Remarks} \label{sec6}

Let us conclude this work by acknowledging that the arbitrage-free pricing and hedging of SOFR caps, swaptions and other derivatives in a multi-curve framework are important practical issues and thus they require further theoretical and numerical studies in a multi-factor model, also with stochastic spreads between overnight rates. For instance, it would be natural to postulate that the basis $r^e-r^s$ is a stochastic process driven by another Brownian motion. Then, in order to ensure the model completeness, one could assume that $B^s,B^e$ and $\Rsf$ are traded instruments. However, it would be important to ensure that the trading arrangements preclude arbitrage opportunities between $B^s$ and $B^e$ relying on simultaneous borrowing cash at a lower rate and lending cash at a higher rate. This in turn would imply that the market model is no longer frictionless and one would need to deal with a nonlinear pricing paradigm, as studied, for instance, in \cite{NR2018} (see also \cite{BCR2018,BR2015,BBFPR2022,BP2014}). In that case, the arbitrage prices and hedge ratios for a SOFR swap and other SOFR derivatives could be obtained by numerically solving a suitable backward stochastic differential equation with a nonlinear generator.


\end{document}